\documentclass{aa}
\usepackage{txfonts}
\usepackage{natbib}{}
\usepackage{amsmath}
\bibpunct{(}{)}{;}{a}{}{,}
\usepackage{graphicx}
\usepackage{placeins}
\usepackage{pifont}

\usepackage{hyperref}

\newcommand{\teff}{$T_{\rm{eff}}$}

\newcommand{\lL}{\ifmmode \log \frac{L}{L_{\sun}} \else $\log \frac{L}{L_{\sun}}$\fi}
\newcommand{\mdot}{$\dot{M}$}

\newcommand{\vinf}{$V_{\infty}$}

\newcommand{\kms}{km~s$^{-1}$}
\newcommand{\rsun}{R$_{\sun}$}
\newcommand{\zsun}{Z$_{\sun}$}
\newcommand{\lya}{Ly$\alpha$}
\newcommand{\ha}{H$\alpha$}
\newcommand{\hb}{H$\beta$}
\newcommand{\hg}{H$\gamma$}
\newcommand{\mum}{$\mu$m}

\newcommand{\msun}{\ifmmode {\rm M}_{\sun} \else M$_{\sun}$ \fi}
\newcommand{\lsun}{\ifmmode {\rm L}_{\sun} \else L$_{\sun}$ \fi}
\usepackage{xcolor}

\def\Oi{[O~{\sc i}] $\lambda$6300}

\def\Oiii{[O~{\sc iii}] $\lambda\lambda$4959,5007}

\def\Nii{[N~{\sc ii}] $\lambda$6584}
\def\Sii{[S~{\sc ii}] $\lambda\lambda$6717,6731}

\begin{document}

\title{Non-LTE atmosphere models of very luminous sources and their applicability to Little Red Dots, quasi-stars, and similar objects}
\author{Fabrice Martins\inst{1}  
 \and Daniel Schaerer\inst{2,3}
 \and Rui Marques-Chaves\inst{2}
}
\institute{
LUPM, Univ. Montpellier, CNRS, Montpellier, France  \\
           \email{fabrice.martins@umontpellier.fr}
\and
Department of Astronomy, University of Geneva, Chemin Pegasi 51, 1290, Versoix, Switzerland
\and
CNRS, IRAP, 14 Avenue E. Belin, 31400, Toulouse, France
}

\offprints{Fabrice Martins\\ \email{fabrice.martins@umontpellier.fr}}

\date{Received / Accepted }

\abstract
{The spectral properties of Little Red Dots (LRDs) differ from those of active galactic nuclei. LRDs may be the first stage of supermassive black holes formation, where the central engine is hidden in a dense gas reservoir, being de-facto quasi-stars.}
{We investigate whether atmosphere models traditionally used for massive stars with strong winds can produce synthetic spectra morphologically similar to those of LRDs.  }
{We compute atmosphere models and synthetic spectra with the code CMFGEN. The models assume a thermalized radiation field at the inner boundary, parameterized by a temperature varying between 5000 and 12000~K. We adopt a typical luminosity of 10$^{10}$~\lsun. The models are spherical, assume an expanding atmosphere, and are computed under non-LTE conditions and for several metallicities. }
{The spectral energy distribution (SED) is different from a blackbody, with a blue optical spectrum. Broad hydrogen emission lines are produced, their wings being formed by electron scattering. The SED near the Balmer and Paschen limit is rather continuous. A Balmer break is predicted for the coolest temperature models provided the wind density is reduced. The SED and Balmer decrement of most LRDs is reproduced by the models, provided they are dust-attenuated with Av$\sim$1.9-2.7. Assuming the absorbed luminosity is re-radiated in the infrared, the energy output at these wavelengths is consistent with observational constraints. The models predict \ion{Fe}{ii}, oxygen and calcium lines. \ion{O}{i} lines at 8446~\AA\ and 1.129~\mum\ are produced mostly by Ly$\beta$ fluorescence. The strength of emission lines from metals depends on input temperature, metallicity, and details of the radiative transfer models. }  
{CMFGEN atmosphere models predict a large number of spectral properties observed in many LRDs. They struggle to simultaneously produce a genuine Balmer break and strong emission lines.
Whether they are more relevant or not to explain LRDs' spectra compared to alternative models is unclear, leaving open the question of the physical conditions in LRDs.}


\authorrunning{F. Martins}
\titlerunning{Synthetic spectroscopy of LRDs}

\maketitle

\section{Introduction}
\label{s_intro}

The physical nature of Little Red Dots (LRD) discovered by JWST \citep{matthee24} remains elusive. They are characterized by a point source morphology, a red optical spectrum associated with a blue ultraviolet (UV) component, producing a typical V-shape spectral morphology \citep{setton25,deugenio25,degraaff25a}, and very often broad emission lines \citep{akins25,hviding25,degraaff25a} among which \ha\ is particularly remarkable. Mostly found at redshifts beyond 3 \citep{akins25,hviding25}, a few examples have nevertheless been detected locally \citep{lin25,ji25}. For these latter cases, as well as a couple of bright LRDs \citep{kokorev25,lambrides25}, high resolution spectroscopy reveals a rich spectrum of emission and absorption metallic lines including \ion{O}{i}, \ion{Fe}{ii}, \ion{Ca}{ii}, \ion{Na}{i}, \ion{N}{ii}. Within this global picture, LRDs tend to show spectral diversity \citep{barro25,pg26,billand26} with important variations in the morphology of the V-shape, the strength of broad lines and the presence or not of specific metallic lines. 

LRDs were initially interpreted as dust-obscured AGNs or old stellar populations, but these scenarios quickly faced major limitations. The absence of X-ray detection, weak or even no spectral variability, and upper limits on the mid infrared emission are not consistent with the presence of type I AGNs \citep{yue24,kokubo25,burke25,maiolino25,zhang25,setton25b,wang25,liu26b}. In addition the black hole (BH) masses inferred from the \ha\ width are orders of magnitudes larger than expected from classical relations between stellar and BH masses in galaxies \citep{maiolino24,jones25}. And the stellar masses required to explain the observed optical and infrared fluxes exceed what is conceivable for mass assembly in the early Universe \citep{labbe24,akins25}. 
\looseness=-1

An alternative explanation for these newly discovered, intriguing astrophysical objects is that they may be supermassive black-holes enshrouded in a dense gas envelope \citep{kido25,naidu25,inayoshi25,degraaff25b}. As such they are also referred to as BH stars and are described by the quasi-stars model presented by \citet{beg08}. In the quasi-star model a BH inside a dense envelope grows by accretion at a high rate and radiates energy at a luminosity corresponding to that of the BH plus envelope mass \citep[see also][]{roman26}. Refinements of this model by \citet{begdex26} predict that several of the observational properties of LRDs are qualitatively reproduced. In particular the spectral energy distribution (SED) shape would be due to the low temperature of the ionized envelope. A high density of free electrons would cause scattering that could account for the broad Balmer lines \citep{rusakov25,chang25}. The high gas density would absorb all X-ray photons. 

To describe the shape of LRD spectra at $\lambda \ga 4000$ \AA\ several studies relied on blackbodies with a temperature of $\sim$4000-7000~K, and modifications of the blackbody \citep{degraaff25b}. To first order, this reproduces the general shape of the SED, with a peak in the optical and a drop at shorter wavelength, mimicking the red part of the V-shape SED of LRDs. The rest-UV emission is attributed to an independent stellar population from the host galaxy, as also suggested my numerous other studies \citep[e.g.][]{inayoshi25,kido25,sun26}. 
Alternatively, \citet{santarelli25} 
used stellar models for evolved stars. They show that the continuum can be broadly recovered by models with low gravity and effective temperatures typical of F- and G-type stars, as also suggested by \cite{ji25}.
\citet{liu26} reach similar conclusions with a similar type of stellar models. In addition they produce absorption lines from calcium that are qualitatively consistent with those observed in some LRDs. The stellar models of \citet{santarelli25} and \citet{liu26} adopt a plane-parallel geometry and LTE conditions. 
\citet{sneppen26} dropped these assumptions in their "cocoon" models which are based on a dynamical and extended envelope, illuminated by a blackbody with $T \sim 10^5$ K to describe emission from accretion onto a central object. They show that under certain conditions on the neutral hydrogen density and column density a Balmer break is observed together with broad hydrogen emission lines.
Other models investigated the reprocessing of an incident SED characteristic of active galactic nuclei (AGN) through a slab of dense gas \citep{inayoshimaiolino25,naidu25,torralba25,ji25b,pacucci26}.
The photoionization code CLOUDY \citep{cloudy} is used and provides the transmitted spectrum.  The input AGN spectrum is either an analytical model or an empirical spectrum. These models are able to explain the optical spectrum of LRDs, especially the Balmer break-like feature, by hydrogen atom collisionally excited in their n=2 level, providing a large opacity in the Balmer continuum. No or little dust absorption is required, and emission lines are predicted. Scattering is generally invoked to explain the width of \ha\ \citep{ji25b}. \ion{Fe}{ii} lines are formed in the outer layers of the covering slab \citep{torralba25}. The UV continuum is attributed to either the un-attenuated scattered emission from the AGN or light from the host galaxy. 

In the present study we compute complementary synthetic spectra that incorporate some of the previous physics as well as new ingredients. In particular we investigate if it is possible to produce a spectrum with 1) the red-optical SED, 2) broad hydrogen lines, 3) helium emission lines, 4) metallic absorption and/or emission lines. We use a non-LTE atmosphere code which is widely used to investigate massive stars' atmospheres, for the reasons that we describe in Sect.~\ref{s_method}. A major difference compared to previous models is that we do not assume any incident spectrum for the source. We consider thermalized emission in a dense atmosphere and self-consistently compute the emergent spectrum. We recall the physical ingredients of the models in Sect.~\ref{s_method}. In Sect.~\ref{s_res} we present the resulting spectra before discussing them in Sect.~\ref{s_disc}. We gather our conclusions in Sect.~\ref{s_conc}.

\section{Method}
\label{s_method}

\subsection{Motivation}
\label{s_etacar}

\begin{figure}[]
\centering
\includegraphics[width=0.49\textwidth]{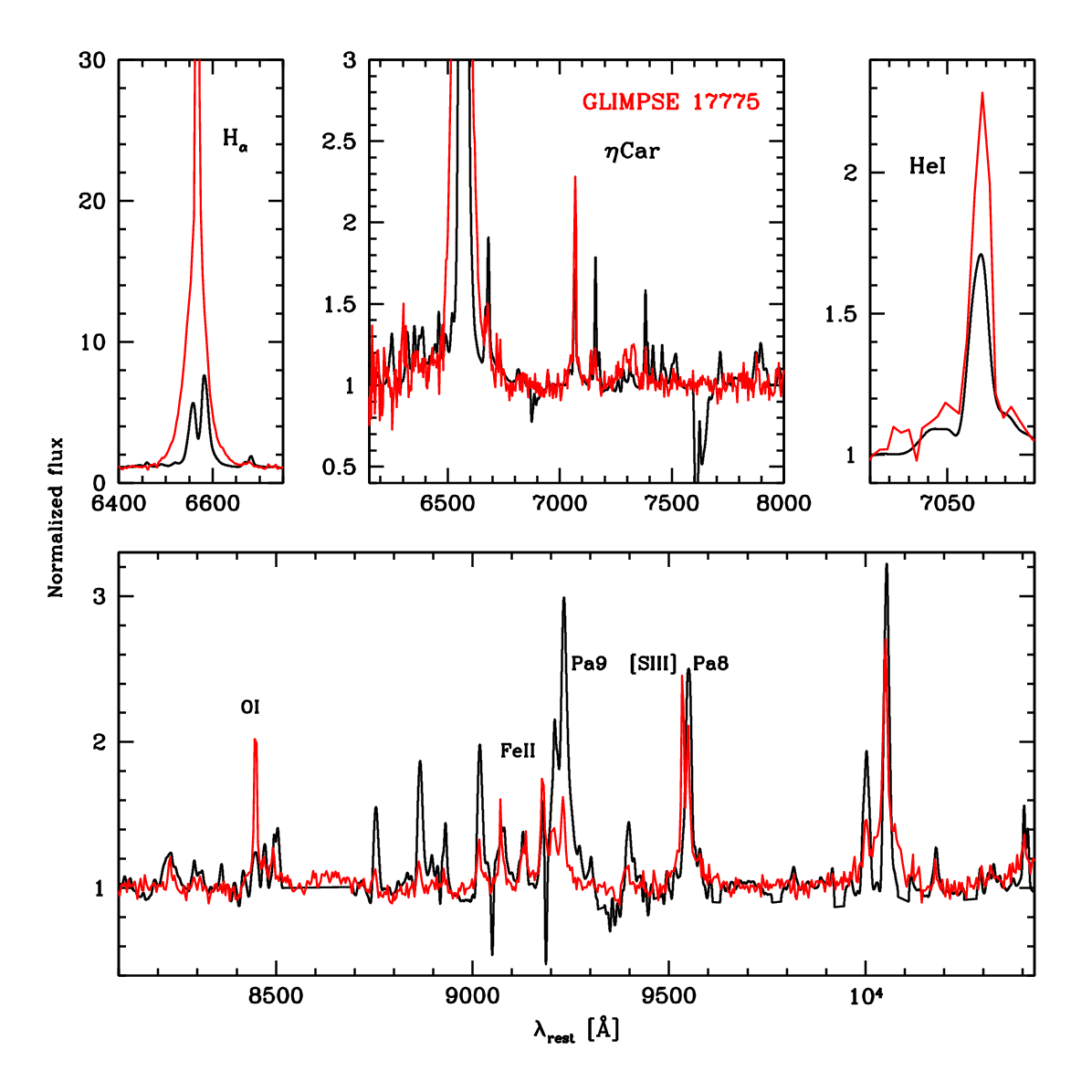}
\caption{Comparison between the JWST spectrum of GLIMPSE 17775 at z=3.5 (red) and the ESO/UVES spectrum of $\eta$ Car (black).}
\label{fig_etacar}
\end{figure}

A few LRDs have been observed at spectral resolution and signal-to-noise ratio that allows the identification of several emission and absorption lines from metals, as well as helium. This is the case of three local LRDs presented by \citet{lin25}. \citet{kokorev25} also obtained a good quality spectrum of the LRD GLIMPSE 17775 (z=3.5) owing to its magnification by a gravitational lens. This spectrum\footnote{Retrieved from the DAWN JSWT archive at \url{https://dawn-cph.github.io/dja/}} is shown in Fig.~\ref{fig_etacar} where it is overplotted on top of the spectrum of the massive star $\eta$~Car\footnote{ESO/UVES spectrum retrieved from \url{https://archive.eso.org/scienceportal/home}}. That star is a luminous blue variable (LBV), a class of massive stars that experience strong stellar wind and rapid evolution. $\eta$~Car is one of the most luminous star in the Galaxy \citep{dh97} and is known for its eruptive phase during the XIX$^{th}$ century. At that time it expelled about 40~\msun\ of material in a few decades \citep{morris17}.

$\eta$~Car displays a rich spectrum of emission lines, among which \ha\ is intense and relatively broad. \citet{hillier01} showed that its wings are explained by electron scattering. \ion{He}{i} emission is present, e.g the 7065~\AA\ feature (see top right panel in Fig.~\ref{fig_etacar}), and a rich spectrum of \ion{Fe}{ii} is also observed. Although differences exist, there are also striking similarities between the spectrum of $\eta$~Car and that of GLIMPSE 17775. The \ion{Fe}{ii} lines around 9000-9200~\AA, the morphology of \ion{He}{i}~7065, the broad wings of \ha, all these features indicate that the physical conditions in the atmosphere of $\eta$~Car may be relatively similar to those of LRDs. 

Similarities also exist between the spectra of LRDs and those of type II supernovae \citep{dessart09,matthee26}, supernovae remnants remnants \citep{lee19}, and Be stars \citep{arias06}. 
This prompted us to calculate "atmosphere" models of LRDs using the same tools as those routinely used for the analysis of massive stars' spectra, in particular those of LBVs. This work is an extension of that presented in \citet{martins20} where we studied the spectral properties and detectability of supermassive stars with luminosities $L \sim 10^{7-9}$ \lsun. We now describe how we proceeded.

\subsection{Atmosphere models and input parameters}
\label{s_etacar}

We have used the code CMFGEN \citep{hm98} to build atmosphere models and the associated emergent spectrum. CMFGEN was developed for the modeling of the atmospheres of massive stars and supernovae. It solves the radiative transfer under non-local thermodynamical equilibrium (non-LTE) conditions, assuming a spherically extended and expanding medium, and including many ions with abundances specified by the user. The code is described in Appendix~\ref{ap_cmfgen}.

\begin{table}[t]
\caption{Main parameters of the models at a metallicity of 0.2~\zsun.}
\begin{tabular}{l|lllll}
model ID    & T$_{\rm input}$ &  \lL     &   R$^a$    &   \teff\     &   \mdot\  \\
            &  [K]        &          &  [\rsun] & [K]        &   \msun/yr  \\
\hline
T12         & 12000       &  10      &  23265   &  3322      &  0.20  \\
T9          & 9000        &  10      &  41360   &  3606      &  0.15  \\
T6          & 6000        &  10      &  93061   &  3276      &  0.30  \\
T5          & 5000        &  10      & 134008   &  4504      &  0.60  \\
\end{tabular}
\tablefoot{a: the radius is given at the inner boundary of the model. It is not the effective radius at which \teff\ is defined.}
\label{tab_param}
\end{table}

In the present study we aimed at obtaining synthetic spectra for a few representative cases of LRDs. We thus adopted typical parameters of LRDs as currently known. For the luminosity, we chose a value of $10^{10}$~\lsun. This corresponds to $10^{43.6}$ erg/s which is representative of the bulk of LRDs \citep[e.g.][]{degraaff25b,greene25}, although some objects reach luminosities nearly a factor 100 higher \citep{ji25,kokorev25}. The models assume a thermalized radiation field at the inner boundary, characterized by a temperature that will be refereed as the input temperature (T$_{\rm input}$) in the following. The input SED is thus a blackbody defined by that temperature at the bottom of the atmosphere. The inner radius directly follows from the assumed luminosity and input temperature. We produce models with an input temperature that ranges from 5000 to 12000~K. We will see that this does not correspond to the effective temperature, \teff, of the models, which is here defined as the temperature where the Rosseland mean optical depth $\tau_{\rm Ross}=2/3$.

Our models are spherical, 1D, and monotonically expanding. We adjust the velocity, and consequently the density, at the base of the atmosphere to ensure that the diffusion approximation is verified at depth (i.e. the optical depth is larger than about 100 and a thermalized atmosphere is recovered). Unless stated otherwise the maximum velocity at the outer boundary of the models is set to 200~\kms. The parameters that define the velocity structure (velocity at the inner boundary, terminal velocity, scale height) alter the shape of the velocity, and thus density, structure. The mass loss rate (\mdot) globally shifts the density structure upward or downward. The main parameters of the models discussed in this paper are listed in Table~\ref{tab_param}. 
Other parameters describing the wind velocity and density, possible inhomogeneities, and microturbulence are discussed in Appendix \ref{ap_cmfgen}. We computed models for several metallicities between 0.01~\zsun\ and 0.4~\zsun, scaling the solar composition.

\section{Results}
\label{s_res}

In this Section we first describe the SED of our models and compare them to those of LRDs. We then focus on specific lines predicted by the models, and discuss their physical properties, formation process, and observability.

\subsection{Overall SEDs}
\label{s_sed}

\begin{figure}[]
\centering
\includegraphics[width=0.49\textwidth]{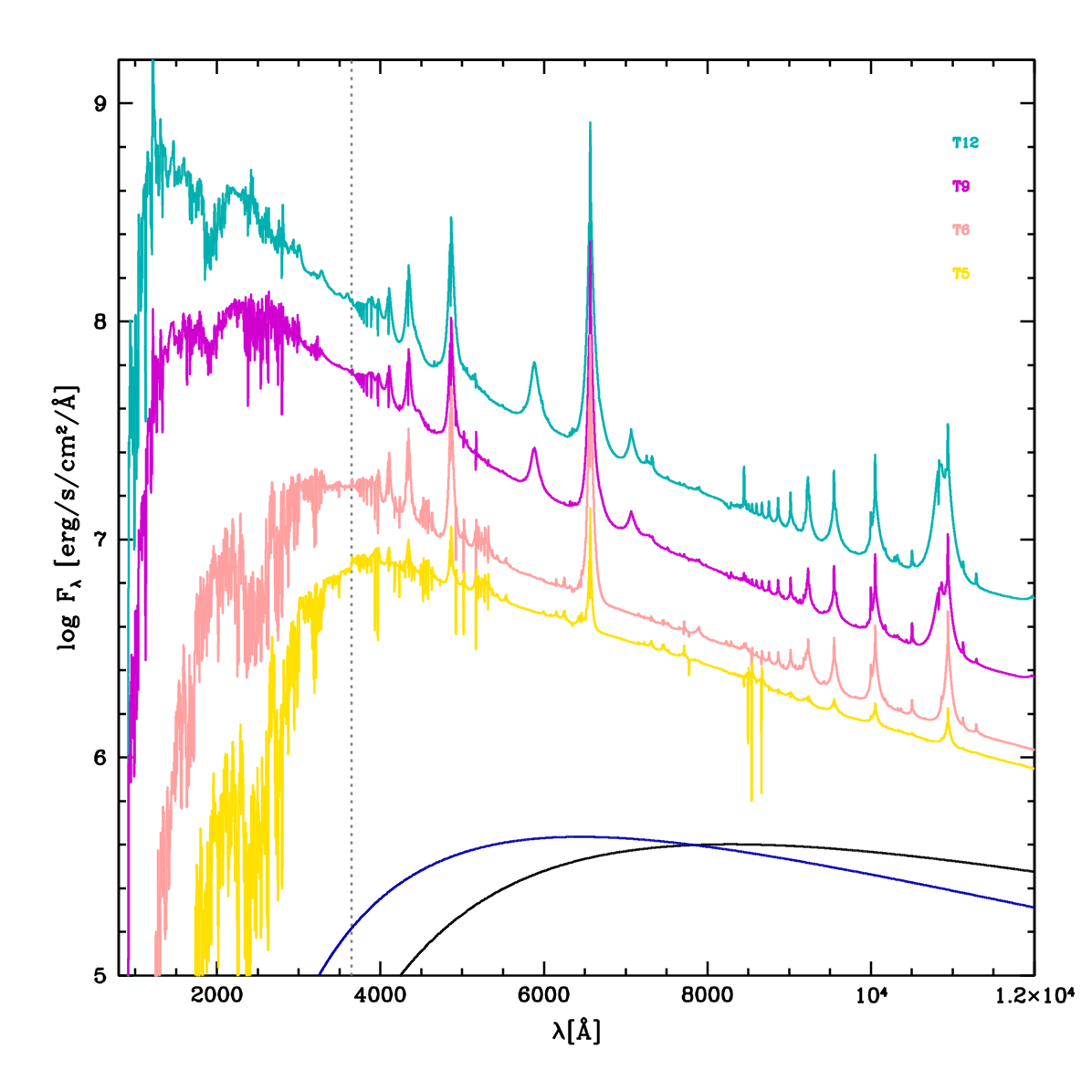}
\caption{SEDs of representative models at Z=0.2~\zsun. The gray dotted lines mark the position of the Balmer discontinuity. The blue and black lines show blackbody SEDs for an effective temperature of 4500~K and 3500~K respectively. The resolution of the spectra is $\sim$1000. 
}
\label{sed_T}
\end{figure}

Fig.~\ref{sed_T} shows the SED of representative models, ordered by input temperature. The SEDs have a maximum that shifts from $\sim$1500~\AA\ to $\sim$4000~\AA\ as the input temperature decreases. For all models \teff\ 
is of the order 3300-4500~K which, as described in appendix \ref{ap_cmfgen}, is different from the input temperature. For such T$_{\rm eff}$ a blackbody SED peaks near 7000-8000~\AA, as seen in Fig.~\ref{sed_T}. 
The model SEDs are thus very different from a blackbody, as already noted by \citet{martins20} for models with similar input physics but lower luminosities, and also reported recently  \citep{liu26,begdex26}. All models show a blue optical continuum and the absence of a strong Balmer break is noticeable. The intrinsic rest-optical SEDs are thus quite different from those of observed LRDs (but see below). 
Many emission lines are predicted, including broad hydrogen lines from the Balmer and Paschen series. Metallic lines are also present and will be discussed in Sect.~\ref{s_lines}.

\begin{figure}[]
\centering
\includegraphics[width=0.49\textwidth]{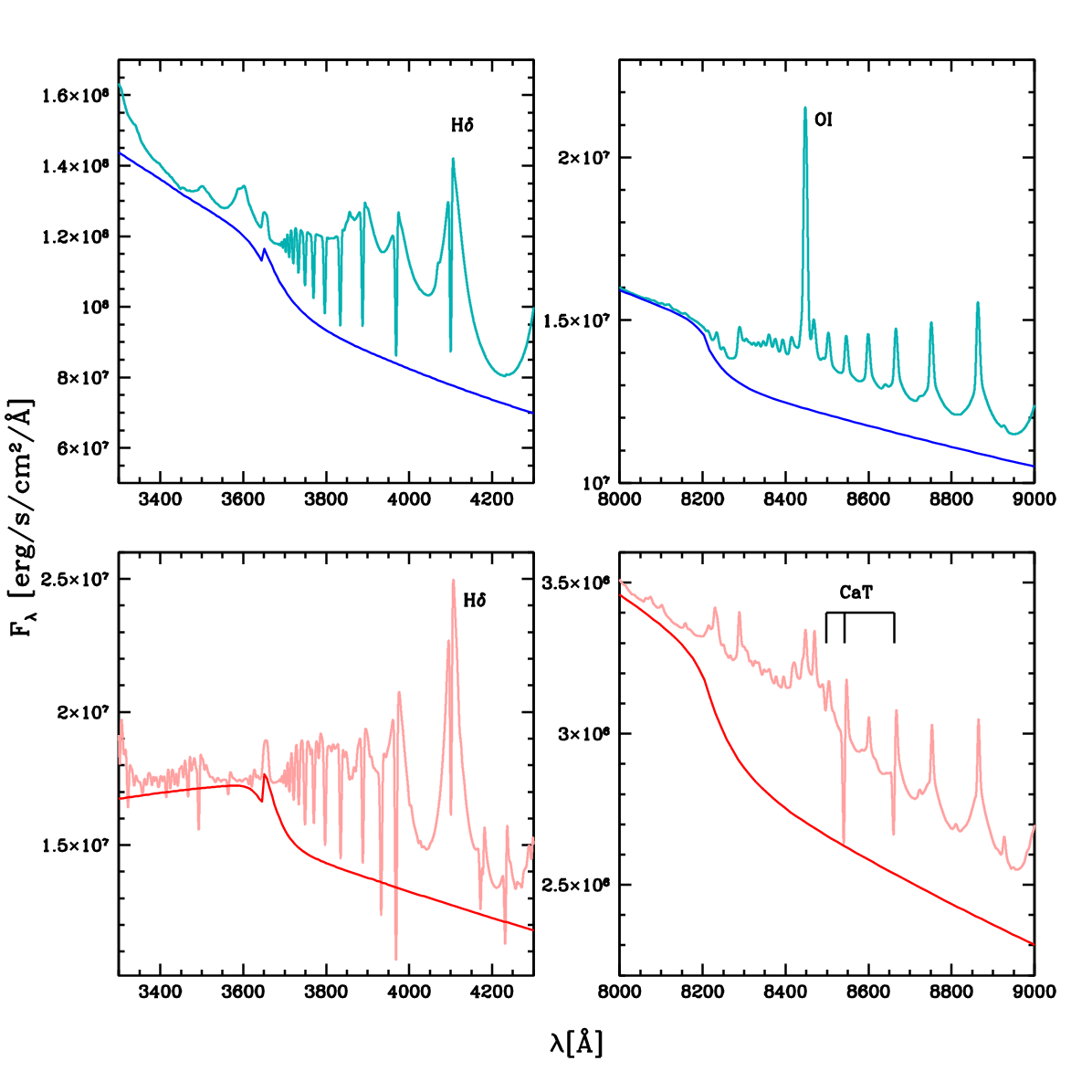}
\caption{Spectra of the T12 (cyan) and T6 (pink) models with their respective continua in blue and red. The left (right) panel shows the region near the Balmer (Paschen) limit.}
\label{sed_T_cont}
\end{figure}

Fig.~\ref{sed_T_cont} helps better understand the nature of the models. The continuum emission is shown for two of the models displayed in Fig.~\ref{sed_T}. Instead of a Balmer break, our models predict a relatively flat continuum or actually a Balmer jump.
The reasons for this behavior have been described in \citet{martins20}. The strong non-LTE conditions in the atmosphere, coupled to the temperature structure, imply a stronger emission below the Balmer break. However, the jump is not clearly seen when lines are taken into account, and a rather continuous flux distribution or a weak Balmer break emerges from the models. This is caused by the broad emission from high-order Balmer lines that tend to overlap near the Balmer limit, creating a pseudo-continuum that smoothly connects to the flux just short of the Balmer break. The SED near the Paschen limit behaves similarly (see Fig.~\ref{sed_T_cont}): the continuum shows a clear jump that is hidden by the high-order Paschen emission lines.

\begin{figure}[]
\centering
\includegraphics[width=0.49\textwidth]{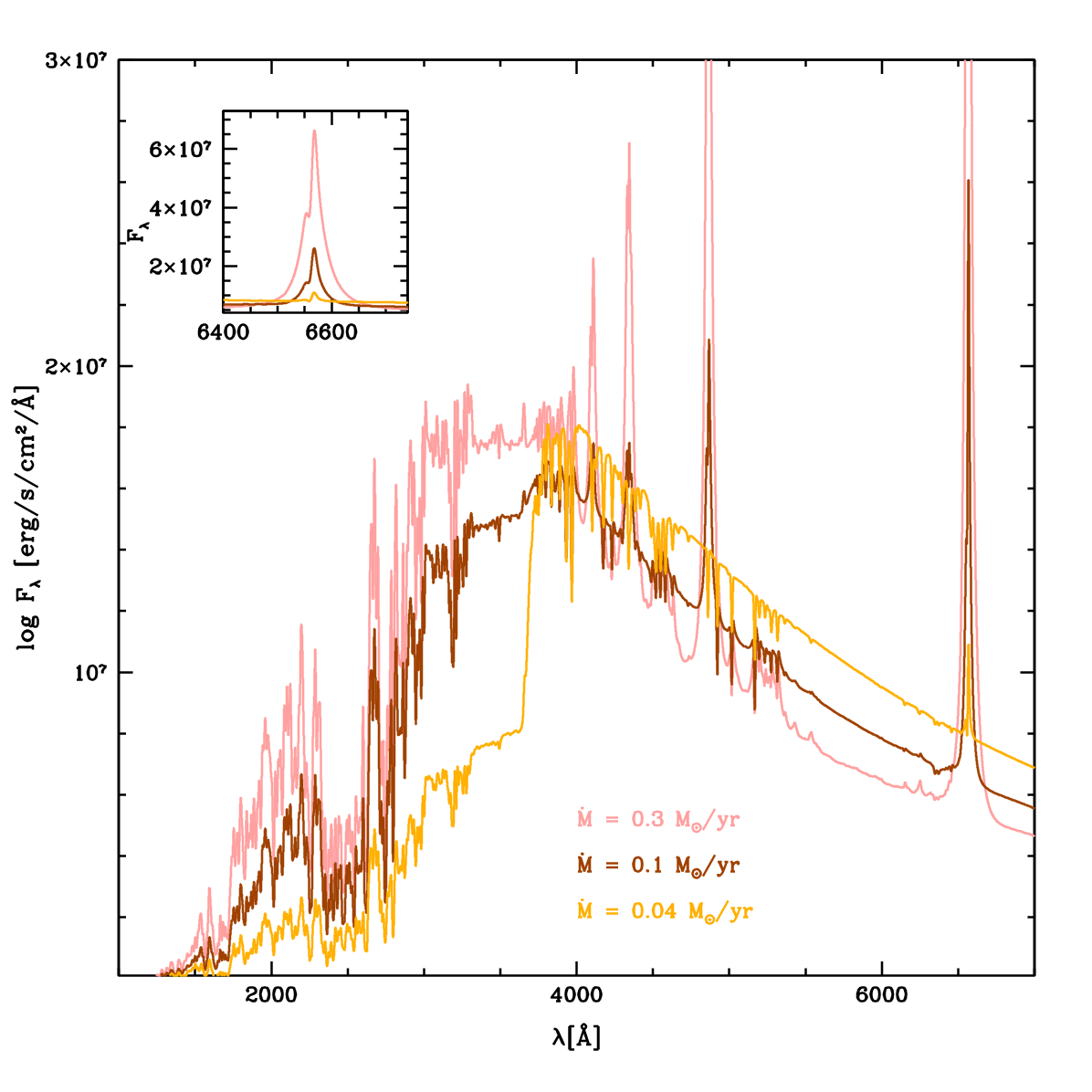}
\caption{Effect of density, modified by varying \mdot, on the emergent spectrum of the T6 model. The spectral resolution is $\sim$500.}
\label{fig_effectMdot}
\end{figure}

\citet{martins20} report that it is possible to produce a Balmer break by changing the atmospheric structure. 
Fig.~\ref{fig_effectMdot} shows that a Balmer break develops in models with reduced density.
At the same time, the strength of all emission lines is severely reduced.  
From top to bottom of Fig.~\ref{fig_effectMdot} the density at an optical depth of unity, representative of the depth at which the optical continuum is emitted, increases. The model with the stronger break has the highest density at that optical depth, while it has the smallest \mdot. This counter-intuitive behavior results from the change in the optical depth scale: in the model with the smallest \mdot, the location of the $\tau=1$ layer is pushed inward compared to models with higher \mdot. Consequently, in the low \mdot\ model the continuum is formed at a position where the density is higher. The ionization structure is also different. Altogether this causes the appearance of a Balmer break.

Independently of the exact density structure, Fig.~\ref{fig_effectMdot} shows that the peak of the SED remains near 4000~\AA, and the optical continuum is blue. If the models represent the physical conditions in LRDs, we thus need to invoke some amount of extinction to reproduce their red optical continuum as we now demonstrate.

\begin{figure*}[]
\centering
\includegraphics[width=0.47\textwidth]{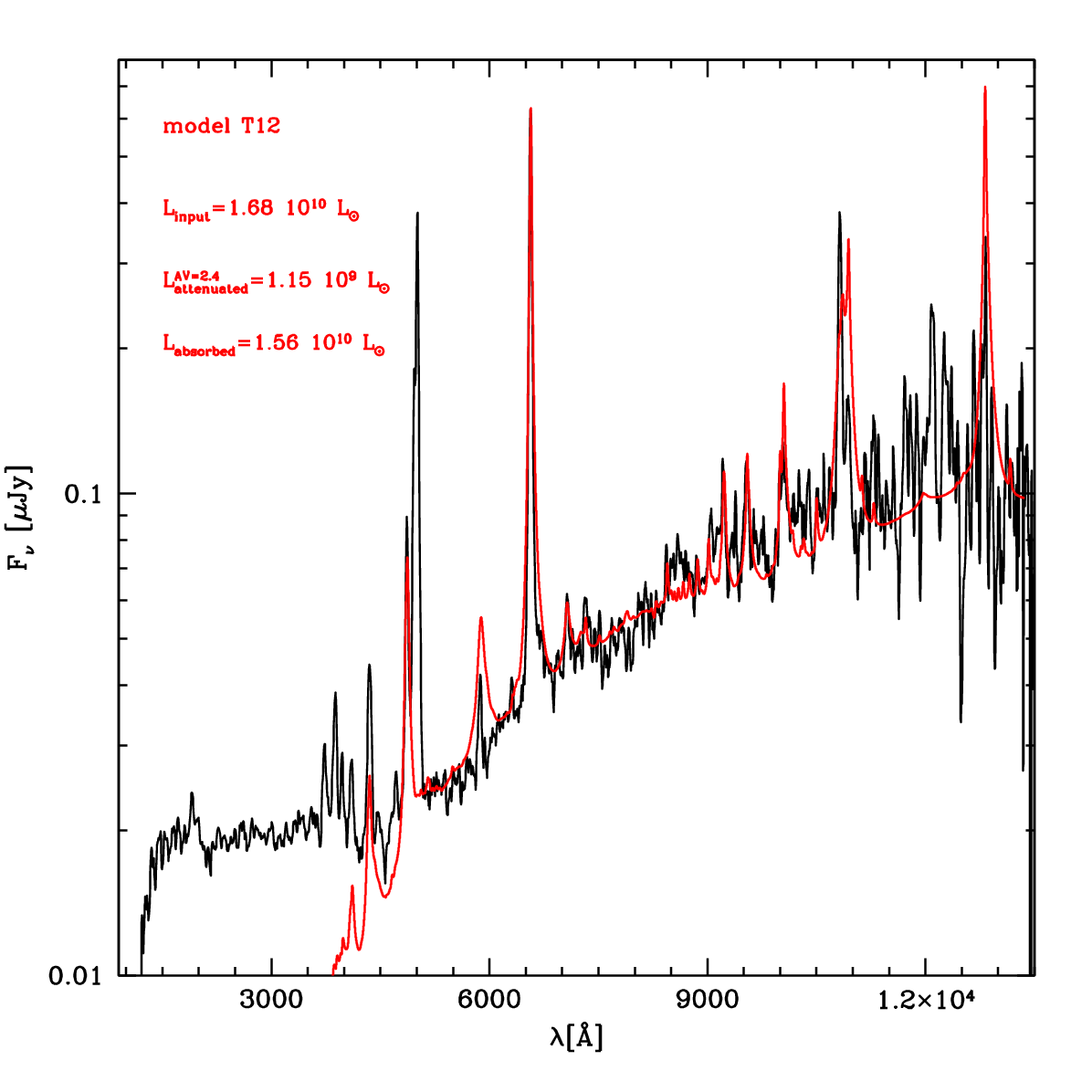}
\includegraphics[width=0.47\textwidth]{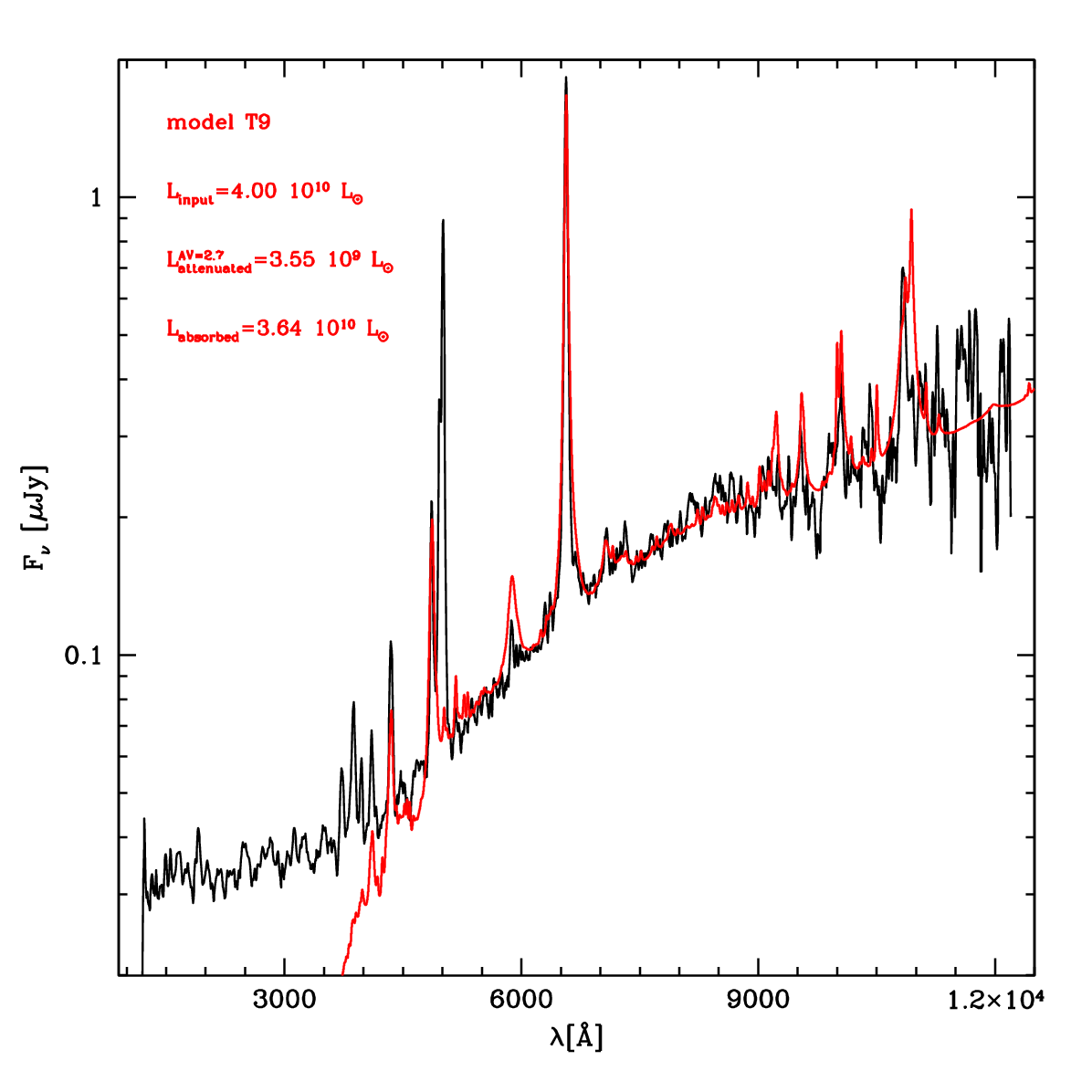}\\
\includegraphics[width=0.47\textwidth]{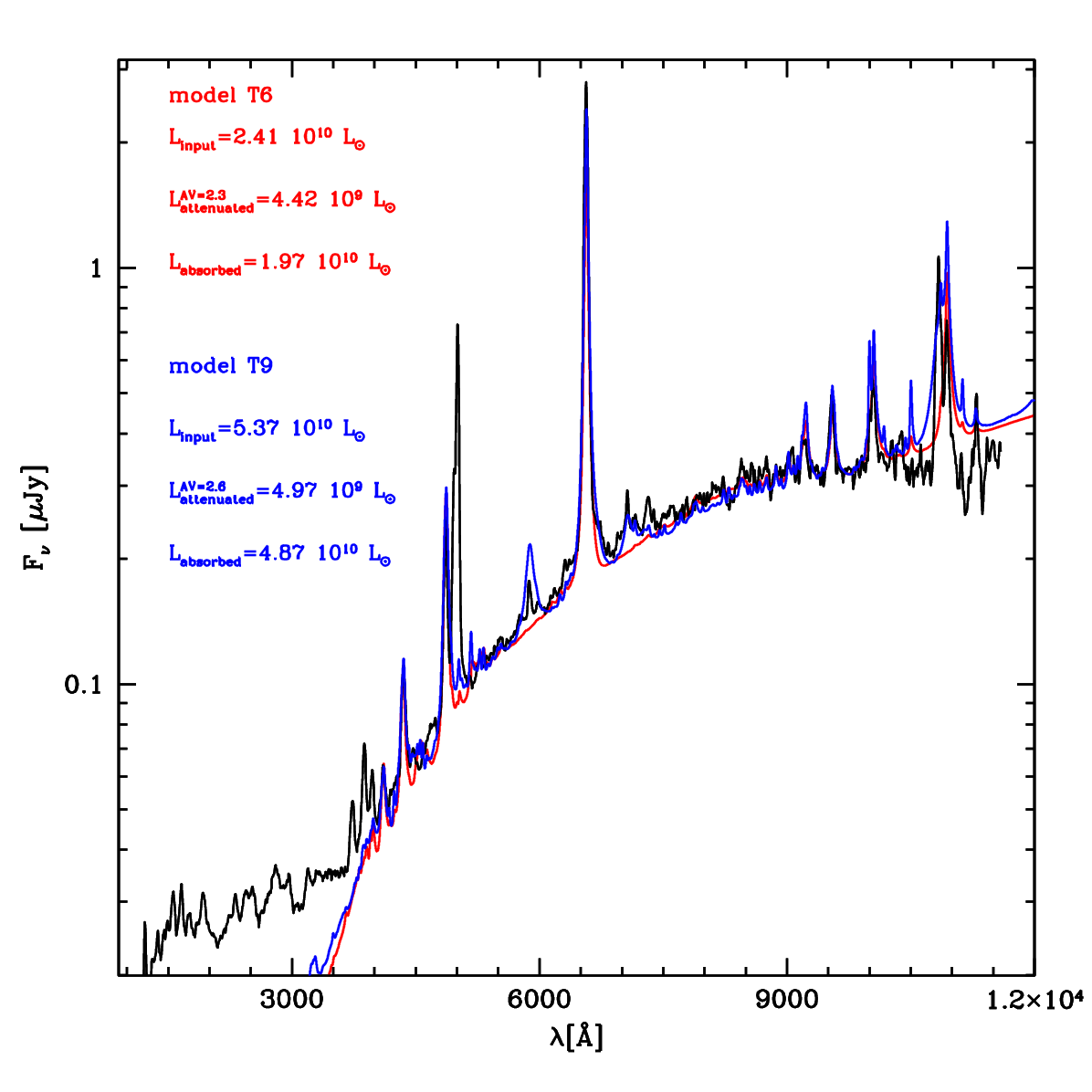}
\includegraphics[width=0.47\textwidth]{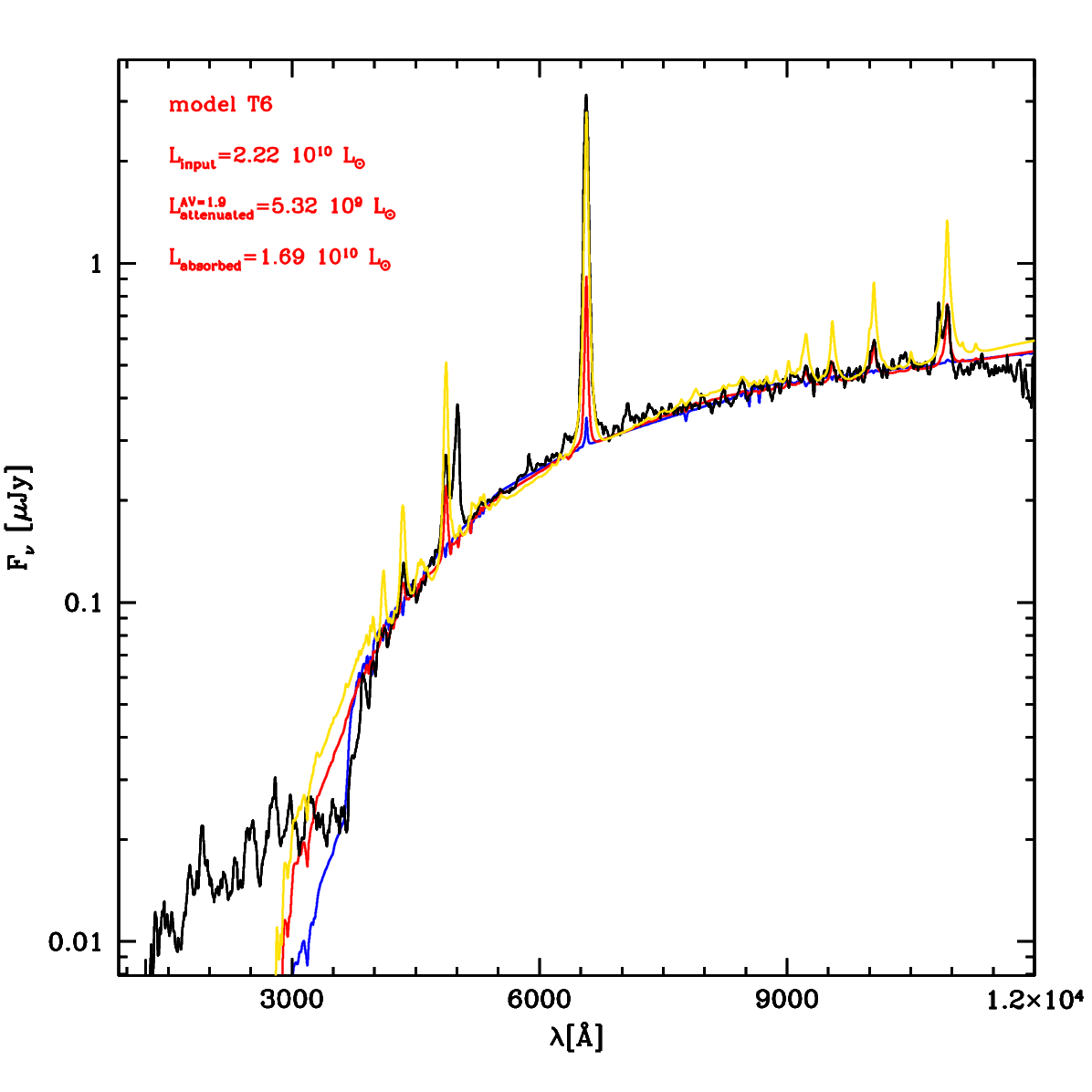}
\caption{SED of selected models in color compared to the stack spectra of \citet{pg26} for bLRDs (top left), -LRDs (top right), +LRDs (bottom left), and xLRDs (bottom right) using Perez-Gonzalez et al.'s nomenclature. In each panel we indicate the intrinsic luminosity of the model (L$_{input}$), the model luminosity after attenuation (L$^{Av}_{attenuated}$), and the difference L$_{absorbed}$=L$_{input}$-L$^{Av}_{attenuated}$. The value of Av used for attenuation is given for each model. In the bottom right panel the yellow (blue) line is a model  with an increased (reduced) mass loss rate. }
\label{sed_pg26}
\end{figure*}

To test the ability of the models to reproduce the observed SED of LRDs, Fig.~\ref{sed_pg26} compares our predictions with observations from stacked LRD spectra covering a range of LRD properties, taken from \citet{pg26}. To reproduce the overall continuum shape we reddened the spectra using the SMC extinction law of \citet{gordon03}. With values of A$_V \sim 1.9 - 2.7$ our models match well the observed SEDs or LRDs.
With the exception of the most extreme LRDs (bottom right panel of Fig.~\ref{sed_pg26}) no true Balmer break is present in the models, only a curvature of the SED caused by extinction. 
We also note that the \ha/\hb\ ratio is fairly well reproduced, which, for our models, is only possible with dust attenuation, as we will discuss in Sect.~\ref{s_lines_h}.
All models reproduce qualitatively the optical continuum down to the minimum of the V-shape. Obviously, a single model is not able to simultaneously explain the UV part nor some narrow emission lines in the optical (e.g., [\ion{O}{iii}]~4959,5007), which we attribute to the host galaxy (not modeled here; see Sect.~\ref{s_disc}). 

To match the different LRD stacks illustrated here with the selected models, the fraction of energy absorbed by dust is typically fairly high, i.e.~$f_{\rm abs} \sim 75-93$ \% of the input luminosity. For observed LRD luminosities which are typically $L_{\rm obs} \sim 10^{10}$ \lsun\ in the rest-optical part of the spectrum \citep{degraaff25b}, this implies that our models would predict dust emission with luminosities $L_{\rm dust} = f_{\rm abs}/(1-f_{\rm abs}) L_{\rm obs}  
\sim (3-13) \times L_{\rm obs} \approx (3-13) \times 10^{10}$ \lsun. These values are compatible with current upper limits on dust emission from \cite{casey25}, who obtained upper limits of $L_{\rm IR} \la 10^{11}$ \lsun\ from stacks of 60 LRDs, and individual upper limits of  $L_{\rm IR} \la 10^{12}$ \lsun\ \citep{setton25b,xiao25}. 
Overall our models are thus able to reproduce the main properties of SEDs of 
LRDs, provided some dust attenuation is invoked. The amount of attenuation required is consistent with current measurements of rest-frame infrared luminosities of LRDs.

\subsection{Spectral lines}
\label{s_lines}

\begin{figure*}[!ht]
\centering
\includegraphics[width=0.99\textwidth]{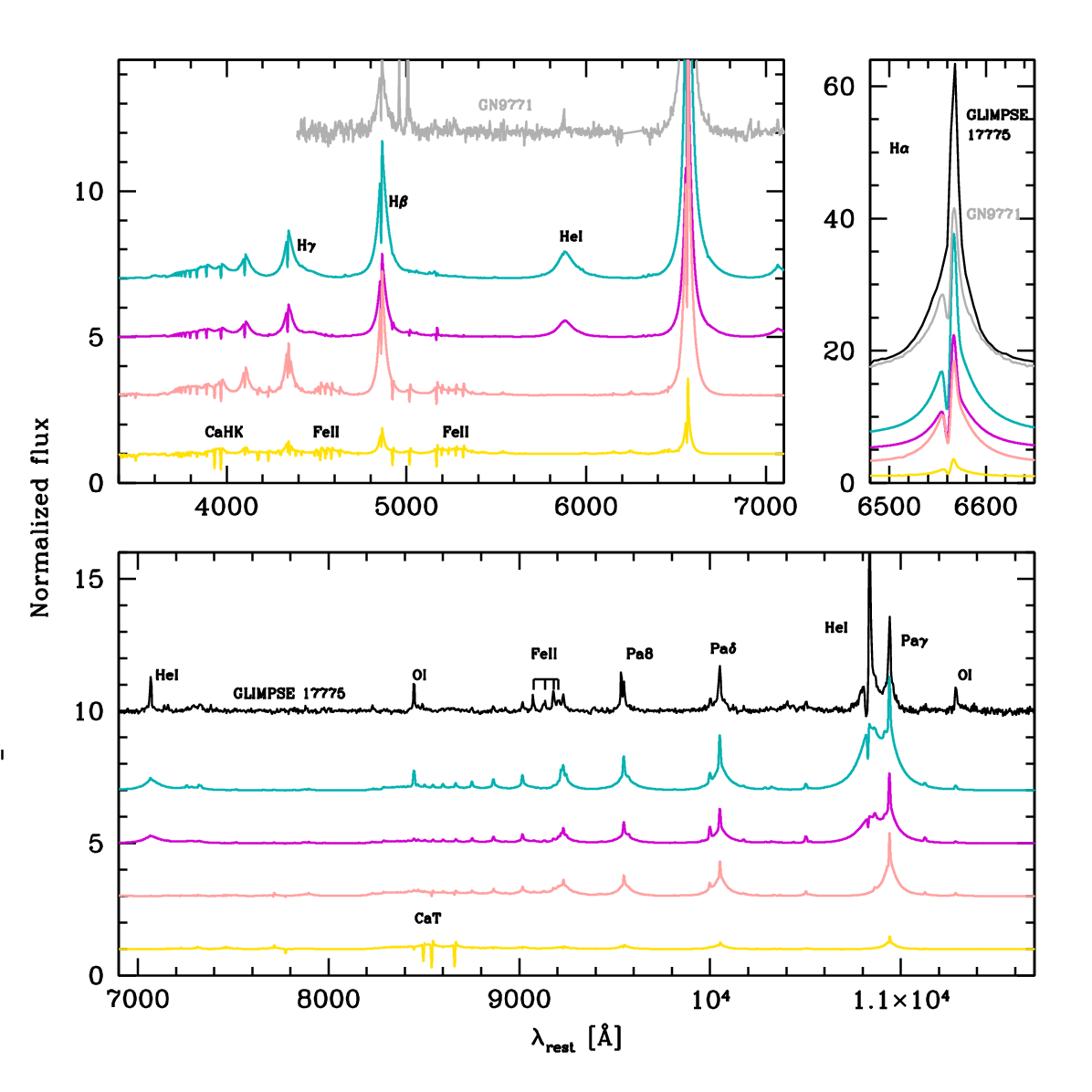}
\caption{Normalized spectra of the models presented in Fig.~\ref{sed_T} compared to the spectra of LRD GN~9771 (gray) and GLIMPSE~17775 (black). Spectra are shifted vertically for clarity.}
\label{normspec_T}
\end{figure*}

We now describe several spectral lines produced by the models.
Fig.~\ref{normspec_T} shows the same four models as in Fig.~\ref{sed_T} now normalized to the continuum. We add the observed spectra of the GN~9771 \citep{torralba25} and GLIMPSE~17775 \citep{kokorev25} to highlight the presence of typical lines observed in LRDs. 
\looseness=-1

\subsubsection{Hydrogen}
\label{s_lines_h}

Broad Balmer and Paschen hydrogen emission lines are produced by the models, with \ha\ standing out as a strong emission feature. 
\citet{rusakov25} argue that the \ha\ profile of LRDs is dominated by electron scattering \citep[see also][]{chang25}. Studying a sample of 32 objects including LRDs \citet{scholtz26} show that the \ha\ profile are not necessarily exponential, as is typical of electron scattering. A significant fraction of objects have profile more consistent with a Lorentzian shape \citep[see also][]{degraaff25a,brazzini25}. These studies favor a stratified medium in which each layer is Doppler broadened at a different velocity, resulting in the observed profile.

\begin{figure}[t]
\centering
\includegraphics[width=0.49\textwidth]{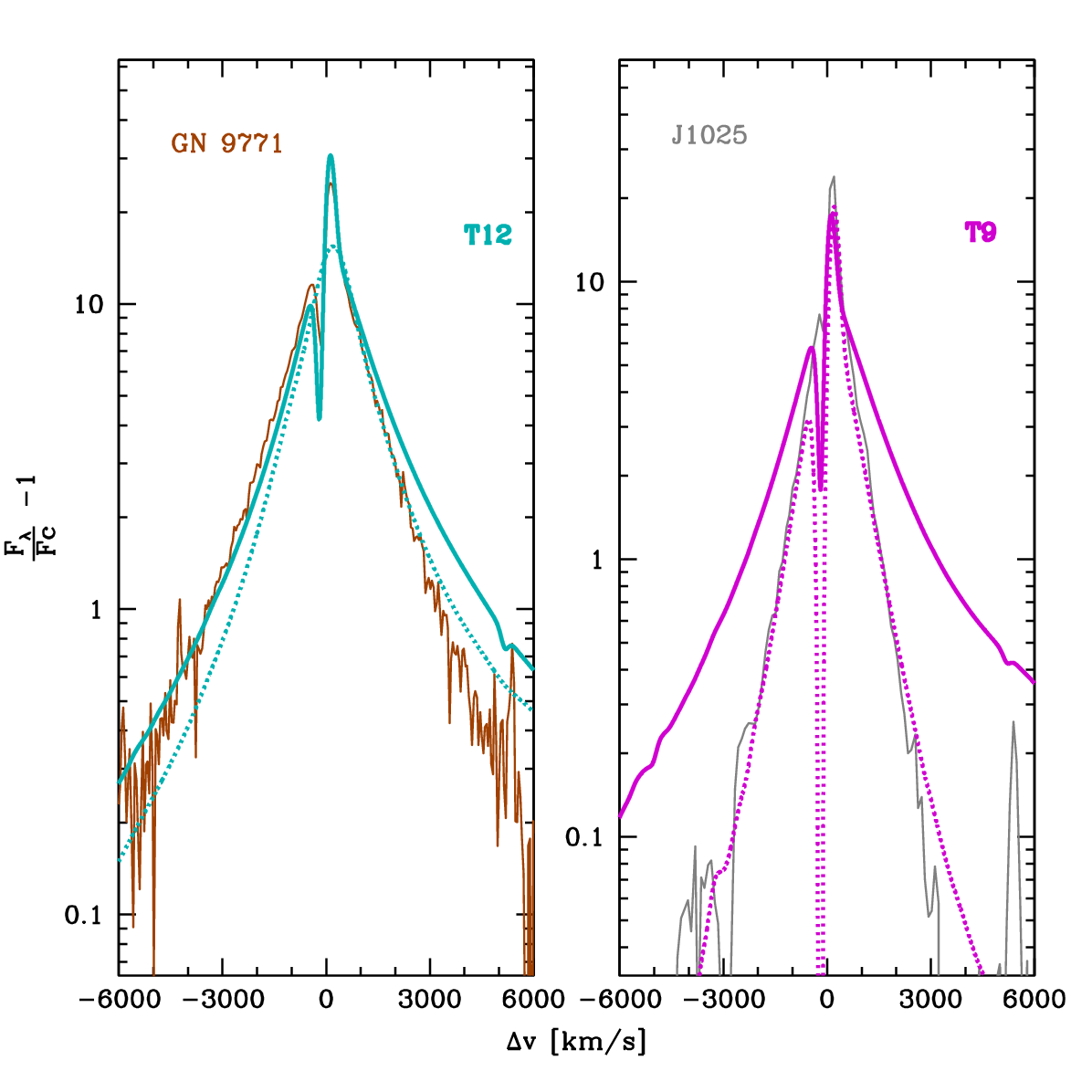}
\caption{\ha\ profiles of two of the models presented in Fig.~\ref{sed_T} together with the profiles of GN-9771 (left panel) and J1025 (right panel). The solid lines are the initial models, the dotted lines models including clumping with a volume filling factor of 0.5 (left) or 0.1 (right). In the right panel the turbulent velocity has been increased to 100~\kms.
}
\label{fig_ha}
\end{figure}

In our models electron scattering is responsible for the broadening of the \ha\ profiles \citep[see also][for a complete description of the process in CMFGEN models]{dessart09}. The extension of the profiles reaches 6000~\kms\ (Fig.~\ref{fig_ha}). Numerical tests limiting the number of scattering produce less extended wings, confirming the physical nature of the process.

The \ha\ emission profile of the models features a blueshifted absorption dip. 
It is naturally produced from a monotonically increasing velocity in a spherical geometry, which creates a P-Cygni profile. 
The absorption part of this profile is almost entirely filled by the emission part, leaving only a weak blueshifted absorption dip. This absorption component is seen in the spectra of some LRDs \citep[e.g.][]{kokorev25,torralba25,brazzini25,scholtz26}. In our case such an absorption dip is a signature of the outflowing nature of the model (see also Sect.~\ref{s_lines_he}).

The T12 model matches relatively well the blue wing, absorption dip and line core of GN-9771 (Fig.~\ref{fig_ha}). However, the red wing exceeds the observed one, since the model \ha\ profile is not symmetric. This is typical of an homogeneous expanding atmosphere in which photons scattered in the receding hemisphere experience more interaction while traveling through the atmosphere to reach the observer. The asymmetry is clearly seen in the right panel of Fig.~\ref{fig_ha}, in the T9 model. The asymmetric profiles are different from those observed in LRDs, where symmetry seems to be the rule \citep[e.g.][]{matthee26}. 
If this is a generic feature of LRD spectra, our models thus need to be adjusted in order to produce symmetric Balmer lines.

A possibility is that the atmosphere is not homogeneous, but clumpy. 
In massive stars, including LBVs such as $\eta$~Car and Wolf-Rayet stars, the electron-scattering wings of hydrogen lines, and in particular the red wing, are sensitive to the degree of inhomogeneities in the atmosphere \citep{hil87,hil91}. 
In addition it is established that massive stars have clumpy winds \citep[e.g.][]{eversberg98,hillier01,bouret05,sundqvist18}. In their analysis of the rest-UV spectrum of the LRD QSO1, \citet{tang26} and \citet{ji26} conclude that an inhomogeneous medium is the best explanation of the morphology of \lya\ and its effects on the formation of UV \ion{O}{i} and \ion{Fe}{ii} lines. 
The dotted lines in Fig.~\ref{fig_ha} show a models in which clumping has been introduced. The details on the formalism can be found in Appendix~\ref{ap_cmfgen}. The effect is striking, especially in the T9 model with a large clumping factor.
Indeed, while the initial profile (solid line) is significantly broader than the observed profile of J1025, the profile of the clumped model almost perfectly matches the observations. To reach this level of agreement we also increased the turbulent velocity in the model from 50 to 100~\kms. This affects the electron scattering wings \citep{hil91} but in the present case most of the difference is caused by the clumpy structure. The increased turbulence causes the deeper absorption dip. 
The fine tuning of the clumped model does not necessarily mean that the physical conditions in J1025 are reproduced by the model. It merely shows that the symmetric \ha\ profile width and intensities comparable to those of some LRDs can be produced, providing the assumption of homogeneity is dropped. Fig.~\ref{fig_ha} also shows that clumping is not always successful in fitting \ha\ profiles. The clumped model in the left panel better reproduces the red wing of GN-9771, but at the cost of a degraded fit of the blue wing. 

Since line asymmetry is caused by the expanding nature of the atmosphere, reducing the expansion velocity produces more symmetric lines. This is illustrated in Fig.~\ref{ap_balsym}. A smaller terminal velocity leads to more symmetric profiles. Line blends and clumping explain that the far wings deviate from symmetry in both panels. The reduction of the expansion velocity also induces a shift of the blue absorption dip towards the line center. 
From these different test models we argue that the asymmetry of the emission lines observed in our initial models can be reduced by adjusting the atmospheric structure.

The models presented in Fig.~\ref{sed_T} have a ratio of \ha\ to \hb\ luminosity of 1.5. This is different from a pure case B nebular ratio ($\sim$3.0), but not unseen in other stellar environments. \citet{briot71} and \citet{banerjee24} report \ha/\hb\ ratios between 1.0 and 2.0 for Be stars, objects known to have a dense equatorial disk. Similarly \citet{levesque14} measure a ratio varying between 1.2 and 3.0 in the LBV star SN2009ip. \citet{chugai04} found that the type IIn supernova SN1994w displayed a \ha/\hb\ ratio in the range 1.0-3.0 depending on the time of observation. Our predicted relative strength of Balmer lines is not specific to the present models, but traces peculiar stellar environments. All are characterized by a dense medium: Be and LBV stars have dense atmospheres, and type IIn are thought to correspond to explosion in a dense layer of material previously ejected by the SN precursor. In App.~\ref{ap_balrat} we demonstrate that this small \ha/\hb\ ratio is likely caused by radiative transfer effects. 

The predicted Balmer line ratios are very different from observations of LRDs showing ratios exceeding case B and up to 10 \citep[e.g.][]{sun26}. Consequently, dust attenuation needs to be invoked to match the observed line ratios.
We have shown in Sect.~\ref{s_sed} that an extinction of Av$\sim$1.9-2.7 was sufficient to reproduce the overall shape of the SED of a large fraction of LRDs. The Balmer decrement is also correctly predicted once dust attenuation is taken into account (see Fig.~\ref{sed_pg26}), except perhaps for the most extreme LRDs (Fig.~\ref{sed_pg26}, bottom right).

\subsubsection{Helium}
\label{s_lines_he}

The spectrum of GLIMPSE~17775 shown in Fig.~\ref{normspec_T} displays clear \ion{He}{i} emission lines \citep[see also RUBIES-BLAGN-1][]{wang25}. The morphology of these features appears to be dual, with a relatively broad component superimposed on a narrow emission core (see below). The models are able to produce broad emission, as is seen in \ion{He}{i}~5876 and \ion{He}{i}~{7065} for instance. The \ion{He}{i}~{1.083}~\mum\ emission is also prominent in the hottest model. The narrow emission is likely due to nebular emission in the host component. The \ion{He}{i} features of the models are quite sensitive to temperature since they completely vanish in the coolest models of Fig.~\ref{normspec_T}. In the three local LRDs presented by \citep{lin25} the one that shows a clear \ion{Ca}{ii} triplet in absorption -- J1025 -- also has the narrowest \ion{He}{i} emission lines, while the the other two objects \ion{He}{i} lines are broader and no \ion{Ca}{ii} triplet is detected. This indicates that a range of ionization conditions is observed in LRDs. Relatively high temperatures are needed to produce \ion{He}{i} emission.

An absorption dip is observed in \ion{He}{i}~{1.083}~\mum\ (see Fig.~\ref{normspec_T}). \citet{wang25} argue that it could be due to an outflow with a velocity of about 200~\kms. Close inspection of this feature in Fig.~\ref{normspec_T} reveals that such an absorption dip is present in the hottest models. Its origin is the same as for \ha: it is part of a P-Cygni profile formed in the expanding atmosphere. 

\subsubsection{Iron}
\label{s_lines_fe}

\begin{figure}[]
\centering
\includegraphics[width=0.49\textwidth]{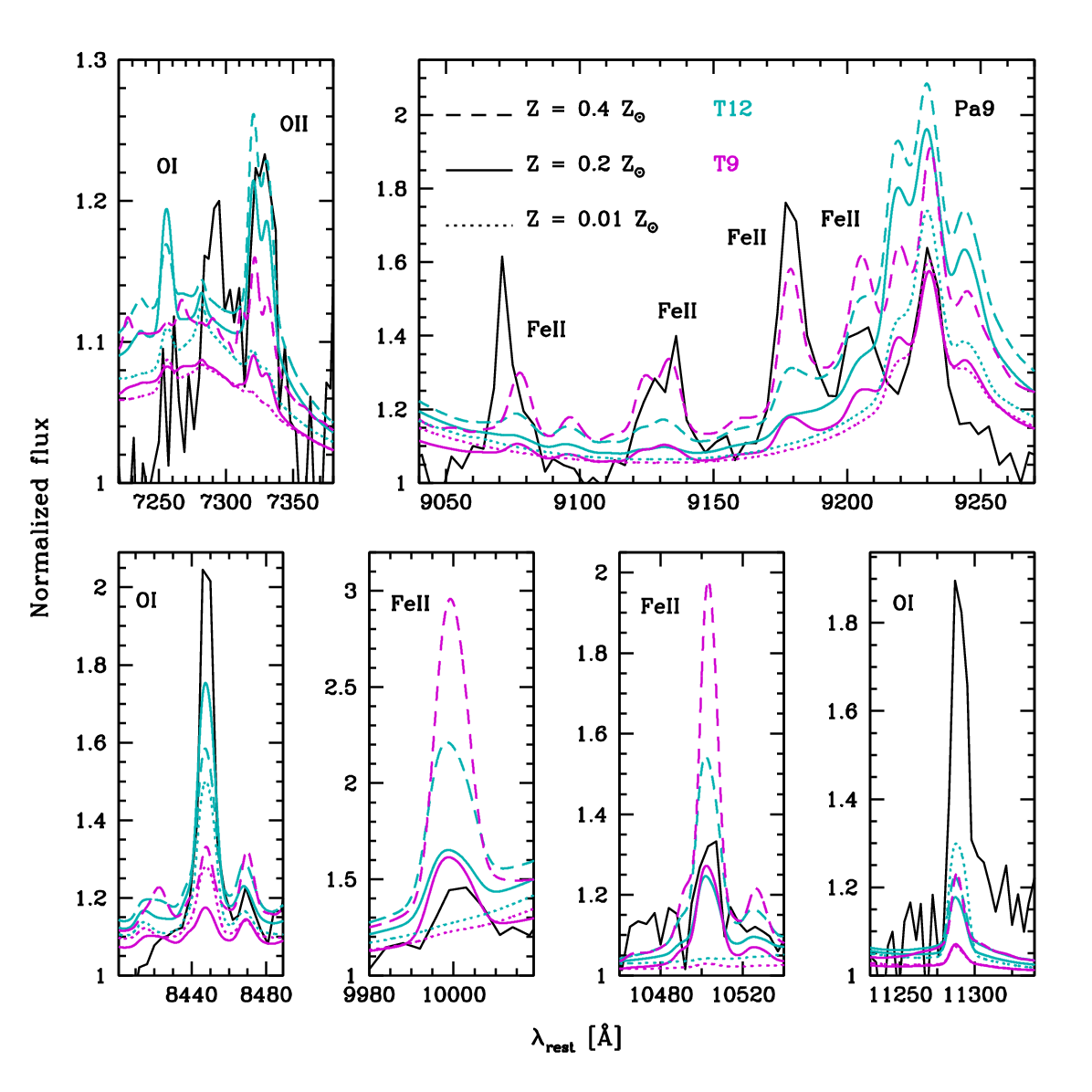}
\caption{Oxygen and \ion{Fe}{ii} lines in the T12 (cyan) and T9 (magenta) models. Solid (dashed, dotted) lines are for Z=0.2 (0.4, 0.01) \zsun. The black solid lines shows the normalized spectrum of GLIMPSE~17775.}
\label{ofe_gl17775}
\end{figure}

\ion{Fe}{ii} lines have been reported in several LRDs. GN~9771 displays several forbidden lines in emission \citep{torralba25}. GLIMPSE~17775 shows emission lines in the near-infrared, as seen in Fig.~\ref{normspec_T}. 
Figure \ref{ofe_gl17775} shows how the hottest models compare to the spectrum of GLIMPSE~17775. Most lines that are observed are also present in the models, with a variable degree of intensity. The \ion{Fe}{ii} lines near 9150~\AA\ are reasonably well accounted for by the Z=0.4~\zsun\ models, especially since the models are not tailored to fit any peculiar LRD. At 9997~\AA\ and 1.049~\mum\  the \ion{Fe}{ii} lines are better reproduced by the 0.2~\zsun\ models. Forbidden lines, e.g. [\ion{Fe}{ii}]~4417, 4453, and 5160~\AA\ are present in our models, especially at the highest metallicities we probed. All lines vanish at low metallicity. Whether these lines can be used for more quantitative metallicity estimates needs to be investigated. \citet{sigut98,sigut03} showed that their formation is complex, involves \lya\ fluorescence mechanisms, and their predicted emission does not perfectly match the observed features of AGNs. \citet{hillier01} also pointed out that the predicted \ion{Fe}{ii} emission spectrum of $\eta$~Car depended heavily on non-LTE modeling, and is affected by uncertainties in atomic physics. 
\looseness=-1

\begin{figure}[]
\centering
\includegraphics[width=0.49\textwidth]{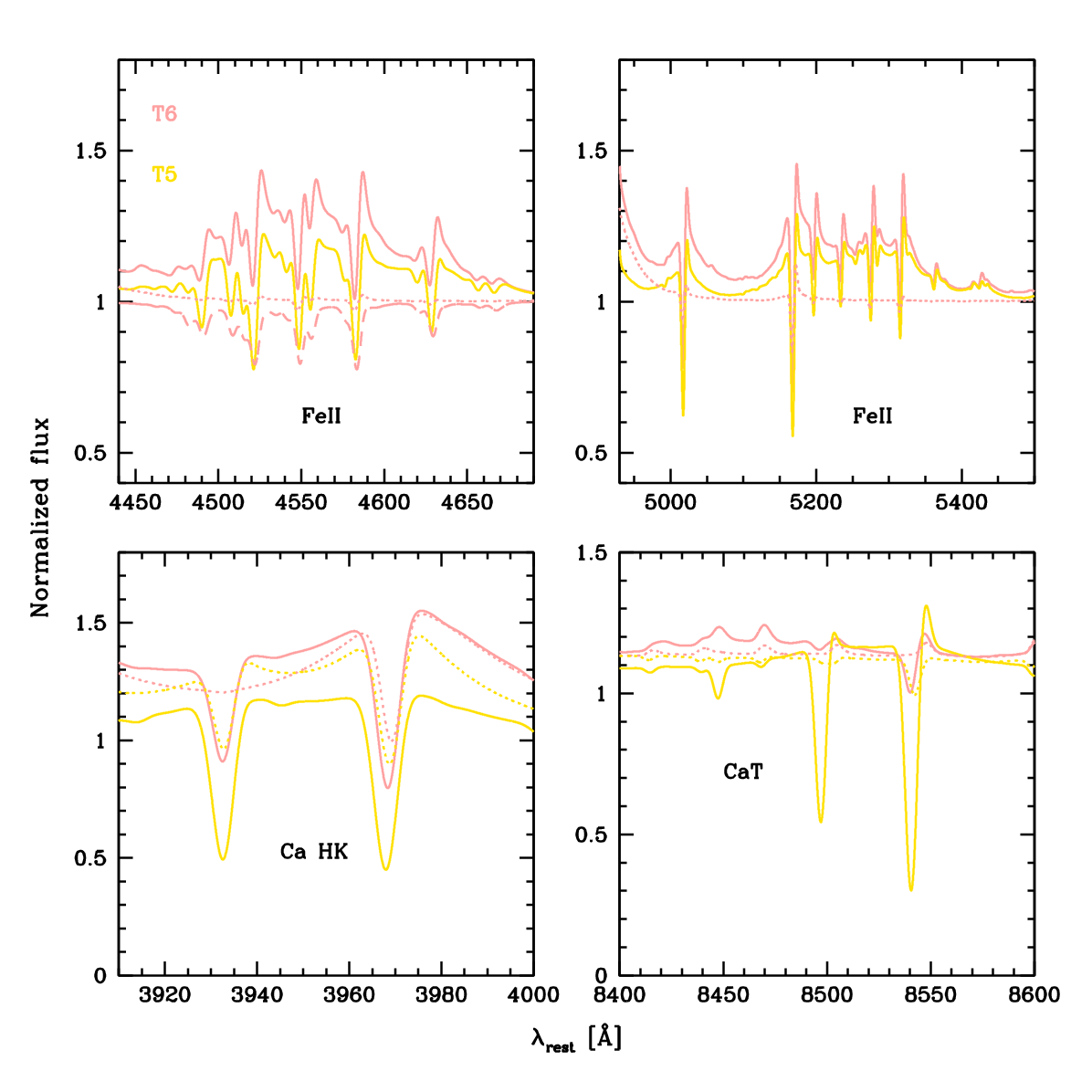}
\caption{Zoom on \ion{Fe}{ii} (top panels) and \ion{Ca}{ii} (bottom panels) lines of T6 (pink) and T5 (gold) models. Solid (dotted) lines are for Z=0.2 (0.01) \zsun. The dashed line in the top left panel is the T6 model at Z=0.2~\zsun\ with the lowest \mdot\ of Fig.~\ref{fig_effectMdot}). The red component of the Ca~H\&K doublet is blended with H$\epsilon$.}
\label{t_L10_CaFe}
\end{figure}

In our models \ion{Fe}{ii} lines are also predicted near 4550~\AA\ and 5250~\AA, but only for the coolest input temperatures. In the T5 and T6 models of Fig.~\ref{normspec_T} they appear as broad emission features within which individual contributions can be seen. Such broad emission complexes are ubiquitous features of AGN spectra \citep{boroson92,marinello16,marinello20,veroncetty04}. Nebular models computed with CLOUDY have reproduced these lines with various degrees of success \citep{panda20,sarkar21,zhang24}. \citet{trefoloni25} shows that the \ion{Fe}{ii} emission near 4550~\AA\ is weak in LRDs and have interpreted this as a sign of low metallicity. 
Figure \ref{t_L10_CaFe} shows that indeed the emission complexes disappear when Z goes significantly below 0.1~\zsun.
However a broad absorption feature near 4500~\AA\ is also reported in some objects \citep[e.g.][]{ji25,pg26}. Figures \ref{fig_effectMdot} and \ref{t_L10_CaFe} show that when the atmosphere density is relatively low the \ion{Fe}{ii} lines around 4550~\AA\ are predicted in absorption, potentially explaining the observed feature provided the metallicity is $\gtrsim$0.1~\zsun.

Our models thus predict different strengths of various \ion{Fe}{ii} lines depending on temperature, atmosphere density, and metallicity. The absence of the emission complex near 4550~\AA\ can be due to a low metal content, but can also be attributed to a relatively hot environment. Constraining the metallicity of the sources thus requires additional information on the physical conditions under which iron lines are observed or not detected.

\subsubsection{Oxygen}
\label{s_lines_o}

The neutral oxygen lines at 8446~\AA\ and 1.129~\mum\ seen in several LRDs are present in the models except the coolest one.
As summarized by \citet{kokorev25} their formation depends on charge exchange reactions and Ly$\beta$ fluorescence \citep[see also][]{tripodi25b}. In Appendix~\ref{ap_OI} we present numerical tests that confirm the role of these physical mechanisms. Predicting the correct intensity is thus complex since this involves sensitive non-LTE radiative transfer calculations. This is illustrated in Fig.~\ref{ofe_gl17775} where the \ion{O}{i}~8446 line does not show a clear trend with metallicity in the T9 model.  This counter-intuitive result stresses that correctly taking into account line opacities near Ly$\beta$ is crucial in the formation and intensity of that \ion{O}{i} line. 

Regarding \ion{O}{i}~1.129~\mum\ the models underestimate the strength of the line observed in GLIMPSE~17775. The line shape is qualitatively reproduced, with an emission peak and an extended red wing, clearly seen in GLIMPSE~17775. 
The models we employ are known to be sensitive to subtle radiative transfer effects when lines of close wavelength compete for photons, as in fluorescence phenomena. \citet{najarro06} reported the dependence of optical \ion{He}{i} lines on the \ion{Fe}{iv} model atom used in the modeling of O-type stars. Similarly \citet{martins12} showed that optical \ion{C}{iii} lines near 4650~\AA\ and at 5696~\AA\ are sensitive to the UV radiation field and to various metallic lines. \citet{rivero11} found the same type of effects for optical \ion{N}{iii}~4634-4642 lines. Consequently it is not surprising to find that the \ion{O}{i} lines discussed above show deviations in strength compared to observed lines. 
\citet{kokorev25} argue that the ratio of the 1.129~\mum\ to 8446~\AA\ \ion{O}{i} lines can be used to measure dust extinction, because their ratio should only be set by atomic physics. This is not the case in our models since in Fig.~\ref{ofe_gl17775} the line ratio varies with metallicity. This most likely results from radiative transfer effects affecting Ly$\beta$ fluorescence as described above.
\looseness=-1

In short, a careful investigation of the \ion{O}{i} formation in the models is necessary before they can be used as quantitative indicators. This is beyond the scope of the present work, and will be pursued in a subsequent study.
In addition to \ion{O}{i}, the models produce the \ion{O}{ii} lines at 7320 and 7330~\AA\ (see Fig.~\ref{ofe_gl17775}). When compared to GLIMPSE~17775 the intensity is qualitatively accounted for. The presence of both \ion{O}{i} and \ion{O}{ii} lines can potentially provide a constraint on the ionization and thus the temperature in the emitting region.

\subsubsection{Calcium}
\label{s_lines_ca}

Fig.~\ref{normspec_T} shows that the coolest models display the \ion{Ca}{ii} triplet in absorption, or at least with an absorption component since the line profile are more of a P-Cygni type. The \ion{Ca}{ii} triplet is observed in some LRDs, for which high resolution spectra have been obtained.
In particular it is detected in J1025, one of the three low redshift LRDs presented by \citet{lin25} and further investigated by \citet{ji25}. 
The calcium H\&K lines are also produced by the models and observed in J1025 \citep{lin25}. In Fig.~\ref{t_L10_CaFe} we see how the calcium lines change with input temperature, metallicity and turbulence. 

Near 8400~\AA\ the presence of \ion{O}{i}~8446 and/or the calcium triplet on top of the Paschen lines can create a broad feature with a step-like structure on the blue side. This is seen in the bottom right panel of Fig.~\ref{sed_T_cont}. A similar morphology has been presented by \citet{lin25} in the spectrum of J1025 (see their Fig.~2).

\section{Discussion}
\label{s_disc}

\subsection{Success, failures, and limitations of our atmosphere models} 

Our models reproduce multiple observational features of LRDs, both qualitatively and sometimes also quantitatively (see summary in Table \ref{t_features}).
However, as for other models, they also require at least one additional component. This is the case to explain the rest-UV continuum emission, narrow nebular \Oiii, and forbidden low-ionization emission lines indicative of low density gas, such as \Oi, \Nii, and \Sii\ seen in some LRDs \citep[e.g.][]{irony25}.
The simplest explanation is that these emission lines originate from the ISM of the LRD host galaxy or its surroundings, and that they are primarily powered by massive star formation in the host or a star cluster surrounding the LRD. This is supported by multiple empirical evidence \citep[e.g.][]{sun26,matthee26}.

\begin{table*}[htp]
\caption{Main observational features of LRDs and ability of models to predict them.}
\begin{center}
\begin{tabular}{l|l|l|l}
Observational feature & CMFGEN  & CLOUDY  & SIROCCO \\
                       & (this work)         &            &     \\
\hline
optical SED shape, Balmer break-like feature        &  \ding{51} dust   & \ding{51} high N(H) & \ding{51} dust, high N(H) \\
Broad \ion{H}{i} lines     &  \ding{51}  electron scattering            &           --$^a$          &  \ding{51} electron scattering \\
\ion{H}{i} line absorption components & \ding{51}                         &  \ding{51}                 & \ding{51} \\
Broad \ion{He}{i} lines & \ding{51}                               &   --$^a$                   &   \ding{51}  \\
\ion{He}{i} line absorption components & \ding{51} & \ding{51}   & \ding{51}  \\
\ion{Fe}{ii} emission    &  \ding{51}                            & \ding{51}                &  --  \\
\ion{Fe}{ii}, \ion{Ca}{ii} absorption  &  \ding{51} at low T and density     &  \ding{51} high N(H)               & -- \\
\ion{O}{i} 8446, \ion{O}{ii} 7320-30 emission  & \ding{51} & -- & -- \\
X-ray / radio weakness &  \ding{51} & \ding{51} & \ding{51} \\
rest-UV continuum & \ding{55}  & \ding{51} , \ding{55}$^b$ & \ding{55}  \\
\protect [\ion{O}{iii}] 4959,5007 emission & \ding{55}  & \ding{55} &  \ding{55}   \\
\protect \Oi, \Nii, \Sii\ emission & \ding{55}  & \ding{55} & \ding{55}   \\
\end{tabular}
\tablefoot{a - CLOUDY models do not include detailed line profiles. b - see text for explanation of the double mark. A \ding{51} (\ding{55}, --) mark indicates that the feature is present (absent, not included or addressed) in the model. Specific conditions to reproduce a given feature are given next to symbols, when relevant. N(H) stands for hydrogen column density.}
\end{center}
\label{t_features}
\end{table*}

Our models assume spherical symmetry and slow outflows, which leads to somewhat asymmetric Balmer lines, in contrast with the observations of some LRDs.
Several studies suggest aspherical symmetries:
For example, the models developed by \citet{sneppen26} can reproduce the symmetry of Balmer lines if both inflows and outflows are considered, and thus multi-dimensional effects are present. \citet{madaumaiolino26} favor a disk-like origin for LRDs, with the diversity in SED shape being due to viewing angles. In that picture there is obviously no spherically symmetric component. And, the observations of a strongly broaded \lya\ line profile in a gravitationally lensed LRD are interpreted by \citet{ji26} and \citet{tang26} as evidence for clumpy structures that may also imply a non-spherical gas geometry. 
However, clumpiness can be treated in one dimensional codes as CMFGEN assuming the clump size is small relative to the characteristic size of the atmosphere. We have shown that clumping may be an alternative solution to explain line asymmetry. 
\looseness=-1

Most studies involving modeling have so far focused on the production of a true Balmer break, i.e. a sharp flux drop at the Balmer limit rather than a smooth roll-over of the SED. As pointed out by \citet{pg26} and \citet{billand26} LRDs that show a sharp and strong Balmer break represent at most 10\% of the LRD population. 
The models we present here are not able to produce both a sharp break and strong Balmer emission lines. Composite spectra can be built out of the models shown in Fig.~\ref{fig_effectMdot}, assuming a fraction of the flux comes from a dense part of the atmosphere (thus producing Balmer emission lines but no proper Balmer break) and the remaining fraction comes from less dense regions (producing a true break but weak Balmer lines). Fine-tuning of the relative fraction of these two components can produce SEDs that show qualitatively both a true Balmer break and Balmer emission lines. If the environment in which LRD spectra are formed is clumped, with strong density variations, such a simplified composite spectrum may apply to describe the most extreme LRDs. 
In any case, the majority of LRDs do not show a genuine Balmer break, but a soft transition from a red optical slope to a blue UV one. The CMFGEN models appear to better reproduce the bulk of LRDs spectra rather than the extreme ones.
\looseness=-1

Despite these limitations our expanding atmosphere models successfully account for the P-Cygni profile of \ha\ and some \ion{He}{i} lines (Sect.~\ref{s_lines}), which can be interpreted as evidence for an outflow. \citet{matthee26} show that LRDs with such line profiles have SEDs with smooth rather than sharp Balmer breaks. They discuss possible origins for these differences. If LRDs are accreting BH with thick equatorial disks, viewing angle may cause variation in the spectral appearance. Alternatively, different stages of evolution, going from a very dense "cocoon" to more optically thin "atmospheres", may be invoked \citep[see also][]{madaumaiolino26}. 
\looseness=-1

\subsection{Comparison to alternative models}

In Sect.~\ref{s_intro} we summarized various attempts to model the spectra of LRDs, beyond adopting (modified) blackbody distributions. They broadly fall under two main categories: studies that use the photo-ionization code CLOUDY and others that rely on stellar-type atmosphere models. To assist the discussion we have summarized features of some of these models in Table \ref{t_features}.

In the first category, a slab of material of constant density is usually assumed. It is illuminated by an incident radiation field and the transmitted radiation flux is predicted. An analytical SED representing a classical type I AGN spectrum \citep{jin12,yue13} is usually assumed as input. \citet{inayoshimaiolino25} show that for gas densities of the order 10$^9$-10$^{11}$~cm$^{-3}$ a Balmer break is produced \citep[see also][]{jimaiolino25}. 
Elaborating on this finding, \citet{naidu25} could reproduce the SED of the LRD MoM-BH*-1 by adjusting slightly the shape of the input spectrum,  gas density and total column density.
A modest extinction (Av=0.15) was also necessary as well as a large turbulent velocity (500~\kms) to reproduce a smooth rollover of the SED instead of a sharp Balmer break.
With a similar approach \citet{torralba25} found a set of CLOUDY models that can produce both a Balmer break and \ion{Fe}{ii} emission lines consistent with the spectrum of GN-9771. Again, gas densities larger than 10$^9$~cm$^{-3}$ are preferred. 
\citet{pacucci26} dropped the assumption of a plane-parallel slab of constant density and adopt the density structure resulting from simulations of direct collapse black holes (DCBH). With a prescribed AGN-like input spectrum for a $10^{5-6}$ \msun\ BH, 
their model reproduces well several main features of the LRD RUBIES-42046, including the Balmer lines and break, and its overall continuum SED from the UV to the optical, with a relatively moderate amount of attenuation ($A_V \sim 0.7$).
While some CLOUDY models can reproduce the overall continuum shape including the rest-UV \citep[see, e.g.,][]{pacucci26}, most of them cannot \citep[e.g.][]{naidu25,inayoshi25}. These latter models generally invoke another origin that is also required to explain forbidden nebular \Oiii\ and low ionization lines since they are suppressed by collisions in high-density environments.
\looseness=-1

A major difference between the CLOUDY models and our atmosphere models is the assumption of an incident spectrum. All models discussed above adopt various forms of type I AGN SEDs, and the emergent  SEDs depend on the input spectra.
In contrast, in our models, the interior of the atmosphere is optically thick at all wavelengths so that the radiation field is thermalized. The input radiation field is thus a blackbody defined by the assumed luminosity and input temperature. 

Similar to CLOUDY models, \citet{sneppen26} assumed a hot black body spectrum ($ T \sim 10^5$ K) as input to their two-dimensional, spherical models computed with the Monte-Carlo radiation transfer code SIROCCO \citep{matthews25}.
In their study \citet{sneppen26} adopted various atmospheric structures but argue that a density $\propto r^{-2}$ gives the best results. Their models produce a variety of SEDs all showing broad hydrogen emission lines (see their Fig.~3) with electron scattering wings. This is consistent with our findings. Most of the SIROCCO SEDs in fact show a blue UV to optical continuum, and only models with high column densities (Ne $\gtrsim$ 10$^{25.3}$ cm$^{-2}$) produce a strong Balmer break. These models reproduce the spectra of LRDs with extreme Balmer breaks, while still producing broad and intense emission lines. This is something our CMFGEN models cannot produce.
The causes of the differences between our models and those of \citet{sneppen26} are not clear. The main differences in the model set-up are the assumed input radiation field, the density structure, and the 1D versus 2D geometry. Radiative transfer is also not computed with the same formalisms (co-moving frame for CMFGEN versus Sobolev approximation
for SIROCCO). \citet{matthews25} performed a comparison between both codes for O-type stars and found reasonable agreement, although also quantitative differences in predicted spectra. 
The only thing that is easily testable is the assumption of the atmospheric structure. A simple test is performed in Appendix~\ref{ap_struc}. The SED shape depends on the density structure but a genuine Balmer break is still not produced in the test model. A more careful investigation of the parameter space is required to see if true Balmer breaks are confined to a specific region. The results of \citet{sneppen26} tend to favor such an explanation.  
\looseness=-1

Other studies, e.g.~\citet{santarelli25}, used LTE plane-parallel stellar atmosphere models computed with ATLAS9 \citep{castellikurucz03}. Their models marginally cover the range of parameters of the quasi-star scenario they explore. The spectra predict a Balmer break, and the Ca H\&K lines are present, although with a too strong absorption. Obviously, since these models are plane parallel, no strong emission lines are predicted. \citet{liu26} used similar plane-parallel models computed with the TLUSTY atmosphere code  \citep{lanzhubeny03}, assuming LTE.
 Focussing on two LRDs, including J1025, they find that low surface gravities, corresponding to low gas densities, are preferred to reproduce the shape of the continuum that significantly departs form a blackbody distribution. They use the calcium triplet feature to estimate the degree of turbulence, favoring values of the order 120~\kms. 
These studies all assume plane-parallel geometry and LTE. The former assumption prevents the formation of broad emission lines, contrary to our models. \citet{martins20} showed that non-LTE conditions prevail in the atmosphere of very luminous and relatively cool atmospheres as the ones presented here. The specific spectroscopic properties of the models described in Sect.~\ref{s_sed} (see also Sect.~\ref{ap_balrat}) confirm that departures from local thermodynamic equilibrium are very large.

\subsection{Implications and future tests}

One of the immediate implications of our optically thick radiation transfer models is that, if applied to a spherically symmetric geometry, the emergent spectrum does not contain any direct observational feature, i.e.~is agnostic of the underlying source powering the base of the atmosphere. In practice, this means that signatures from a possible central BH illuminating the ``cocoon'' of an LRD cannot be observed if the outermost layers are even approximately described by the atmosphere structures adopted here. Quantities such as BH masses can therefore not be inferred with our models.

On the other hand, spherically extended atmosphere models as the ones calculated here depend on the luminosity of the object, which can thus be constrained -- together with other physical properties -- from a comparison between predicted and observed spectra. The comparison with typical LRD spectra \citep[stacks from][]{pg26} has shown that our models reproduce quite well the observations with dust attenuation corresponding to a ``bolometric correction'' BC $=1/(1-f_{\rm abs}) \sim (4-14)$ -- see Sect.~\ref{s_sed}. The intrinsic luminosity of the red part of LRDs (i.e.~excluding the UV component) is therefore $L_{\rm int} \approx {\rm BC} \times L_{\rm obs} \approx (4-14) \times 10^{10}$ \lsun, adopting a typical value of $L_{\rm obs} \approx 10^{10}$ \lsun. 
If we assume that the object is at its Eddington luminosity, the total mass is
\begin{equation}
\frac{M}{\msun} \approx 2.7 \times 10^{-5} \times {\rm BC} \times \frac{L_{\rm obs}}{\lsun} \sim (1-4) \times 10^6 \frac{L_{\rm obs}}{10^{10}\lsun}.
\end{equation}
This mass-luminosity relation is predicted to apply, e.g., to supermassive stars and quasi-stars in hydrostatic equilibrium, independently of metallicity \citep[cf.][]{roman26}. In all cases the mass refers to the total mass, which includes the BH mass plus envelope for quasi-stars, or the total stellar mass for objects powered by nuclear burning.

Detailed non-LTE radiation transfer models, such as the ones shown here,
can in principle also be used to determine the metallicity of LRDs. This would be an important constraint regarding their formation. The DCBH model usually requires metal-free or extremely metal poor conditions to prevent gas fragmentation \citep[e.g.][]{loeb94,ferrara14}, although high pressure metal-rich environments may also form DCBH \citep{mayer24}. DCBH has been advocated as a viable pathway to form LRDs \citep[e.g.][]{jeon26,cenci25}. We have seen that metallicities below 0.1~\zsun\ seemed to be required to avoid the formation of \ion{Fe}{ii} emission bumps around 4550 and 5250~\AA. At the same time, reproducing \ion{Fe}{ii} emission lines near 9150~\AA\ is possible only for metal contents typical of the Magellanic Clouds (0.2-0.4 \zsun). Does this mean that LRDs show a variety of metallicities, in line with their spectroscopic diversity? Answering that question requires a better understanding of metal line formation, and dedicated studies of LRDs for which medium-resolution high signal-to-noise spectra are available. 
This plea applies in general to our models, CLOUDY modeling, and others.

In parallel, further work is needed to test the applicability of our models to LRDs.  
In particular, our models imply that dust must be present in LRDs. \citet{brazzini26,barro26} find that such a component with a temperature of about 1000~K is required to explain the rest-IR measurements of some LRDs, including the so-called Rosetta Stone \citep{juodz25}. From the stack of 60~LRDs \citet{casey25} place upper limits on the IR luminosity, which are broadly consistent with our predictions (see Sect.~\ref{s_sed}).
The need for dust attenuation is a major difference between CMFGEN and CLOUDY models. The latter predict the correct overall SED shape from gas absorption only. The presence or absence of dust in LRDs is thus a key element to test which category of models is the most relevant to describe them. 
\looseness=-1

If LRDs involve supermassive stars at some point of their formation, one may expect to detect nucleosynthesis products in their spectra \citep{roman26,nandal26}. In fact a fraction of LRDs are also classified as N-emitters \citep{tripodi25}, a class of emission-line galaxies that show elevated nitrogen-to-oxygen ratios \citep{cameron23,morel25,naidu26}. Furthermore, supermassive stars are one of the candidates to produce the chemical imprints of N-emitters \citep{charbonnel23,Nagele2023,marques24}. Measurements of the CNO  abundances and other elements in LRDs may thus provide additional insights on the possible link between supermassive stars and LRDs.

\section{Conclusion}
\label{s_conc}

We have computed state-of-the-art, line blanketed, non-LTE atmosphere models and synthetic spectra for very luminous objects with luminosities $\sim 10^{10}$~\lsun, typical of LRDs, quasi-stars and related objects. We have used the  CMFGEN code, traditionally employed for massive stars and supernovae, covering a range of input temperatures and metallicities. The models assume spherical symmetry and an expending geometry in the form of a wind-like structure, and compute the atmosphere and radiation transfer from the optically thick region to outer boundary. The main results are:

\begin{itemize}
    \item The SEDs differ strongly from blackbodies, with a blue optical spectrum and effective temperatures of the order \teff $\sim 3500-4500$ K, much lower than the input temperatures. This is caused by the optical thickness of the models that translates into the position of the photosphere at a radius much larger than the inner radius of the model.

    \item No Balmer break is produced, except in the coolest models that have a reduced atmospheric density. Reproducing the observed SED shape and Balmer line decrements of LRDs requires dust attenuation, with typical values of Av $\sim 1.9-2.7$ The predicted attenuation is compatible with current upper limits on their IR luminosity and individual detections attributed to host dust in LRDs.

    \item Broad Balmer lines are produced in models that have dense atmospheres. Electron scattering dominates the formation of the extended wings. Lines are asymmetric, with more flux in the red wing, as expected of outflowing structures. Symmetry can be recovered if 1) the expansion velocity is a few tens of \kms, or
    2) a clumpy, but optically thick atmosphere is assumed. 

    \item Helium, Oxygen, Iron, and Calcium lines resembling those of LRDs are produced by our models, with intensities that depend on temperature, density, metallicity, and radiative transfer effects. 
    We confirm that the \ion{O}{i}~8446 and 1.129\mum\ lines are produced by Ly$\beta$ fluorescence. In depth studies of the formation process of metallic lines are required before they can be used to infer physical properties of LRDs. 

\end{itemize}

\noindent Compared to alternative models currently available our models differ mostly in that they are optically thick at the inner boundary, and make no assumption relative to the energy source powering LRDs. Whether the assumption of an optically thick environment is appropriate to interpret LRDs and related objects is presently unclear. Further work, including detailed quantitative analysis of high resolution spectra with state-of-the-art models such as those presented here should help clarify this situation and shed more light on the nature of these objects.

\begin{acknowledgements}
We thank John Hillier for making CMFGEN available and for constant help with it.
\end{acknowledgements}

\bibliographystyle{aa}
\bibliography{lrd_atm}




\begin{appendix}

\section{CMFGEN models}
\label{ap_cmfgen}


In this Appendix we describe the main features of the code CMFGEN. The reader should refer to \citet{hm98} for an exhaustive presentation.

CMFGEN requires an input hydrodynamical structure that can have various forms. In the present study we have adopted the classical wind structure of expanding outflows of massive stars. The inner structure is assumed to be in quasi-static equilibrium and is parameterized by a pressure scale-height and a velocity at the base of the atmosphere. It is connected to a wind structure where the velocity is of the form $v = v_{\infty} \times\ (1-R/r)^{\beta}$ where $r$ is the radial coordinate, $R$ the inner atmosphere extension, and \vinf\ the maximum velocity at infinite radius. $\beta$ is a parameter that is typically set to unity. The two structures are connected at a velocity typically chosen around 1~\kms. We adopted \vinf\ = 200~\kms, unless stated otherwise. The density structure naturally follows from this velocity structure and the equation of mass conservation which can be written $\rho = \frac{\dot{M}}{4 \pi r^2 v}$, with $\dot{M}$ the mass loss rate. Clumping can be included under a volume filling factor formalism \citep[see][]{bouret05}. The filling factor $f$ varies with velocity as $f_{\infty} + (1-f_{\infty})e^{-v/v_{cl}}$ where $f_{\infty}$ is the maximum filling factor at the top of the atmosphere (and is what is quoted as the clumping factor in CMFGEN models), $v$ is the velocity, and $v_{cl}$ is the characteristic velocity of the clumping onset.

The inner structure is set so that the conditions of the diffusion approximation are recovered in the inner regions. Radiative transfer calculations are then performed in an iterative way. Level populations of the various elements included in the model are given by the statistical equilibrium equations. Radiative and collisional processes are taken into account, which request the radiation field from the previous iteration to be known. The source functions and opacities are computed from the level populations, and used to predict the radiation field of the next iteration from the solution of the radiative transfer equation. Convergence is achieved when level populations do not vary by more than a control value, typically less than 0.1\%.  In this iterative process, both line and continuum opacities are taken into account. Lines are assumed to have a Gaussian shape, and can be artificially broadened by a turbulence velocity that we fix at 50~\kms. The ions to be included, as well as the number of electronic levels for each ion, can be varied. 

A super-level approach is used to speed-up the calculations \citep{anderson89}. This means that some levels are grouped into a single super-level and share the same departure from LTE. This approach has been validated by decades of atmosphere model computations for massive stars \citep{hm98,hillier01,martins04,bouret13,groh14,besten14,martins24}. 
Because of the non-LTE nature of the code, the interaction of photons with any transition that matches its rest-wavelength is naturally taken into account in the prediction of the level populations. This is usually referred to as "blanketing" effect in stellar atmospheres. The formation and intensity of some lines critically depend on these interactions \citep[e.g.][]{najarro06,rivero11,martins12}. Fluorescence phenomena are a sub-category of these radiative transfer effects and are thus included in the models. 
Charge exchange reactions are also taken into account. A number of ions are treated by this process, in particular \ion{Fe}{ii-iii-iv}. \citet{hillier01} showed that charge exchange is an important ingredient of the formation of \ion{Fe}{ii} emission in the atmosphere of $\eta$~Car. In the models presented here, we included ions from H, He, C, N, O, Mg, Al, S, Si, Ca, Fe. The selection of ions for each elements depended on the input temperature of the models. We tested that H$^{-}$ has no effect in any of our models, its relative abundance being at most 10$^{-5}$, but well below in the vast majority of the models.

Once the atmosphere model is converged, a final solution of the radiative transfer solution is performed in which the level population are held fixed. More realistic line profiles are used, including in particular pressure broadening and a depth variable microturbulence ranging from 10~\kms\ at the base of the model up to 50~\kms\ in the upper layers. In that part of the modeling non coherent electron scattering caused by electron thermal motions is taken into account. We calculate the spectrum between 1000~\AA\ and 2.5~\mum.

\begin{figure}[!ht]
\centering
\includegraphics[width=0.49\textwidth]{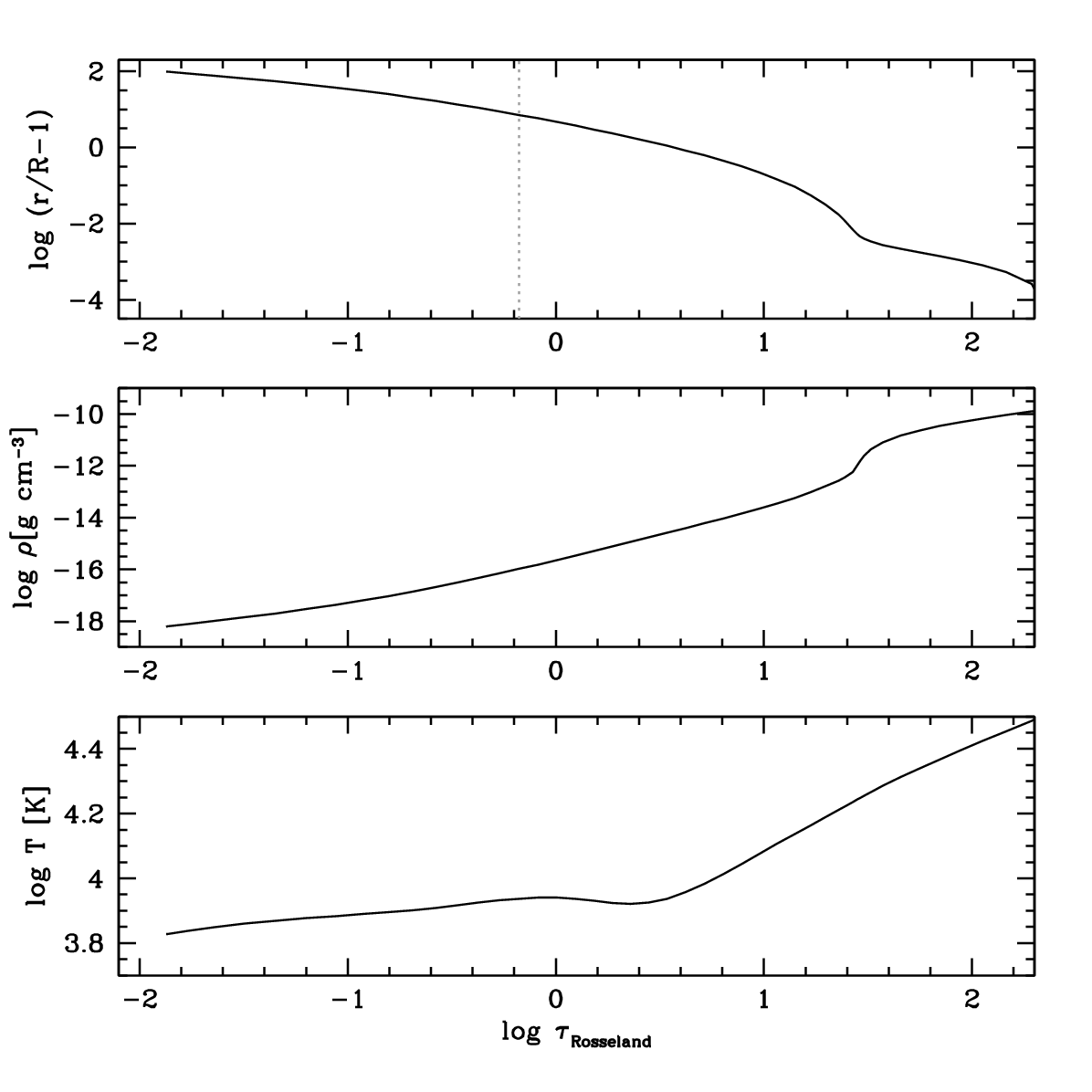}
\caption{Radius, in the form of the height above the bottom of the model (top panel), mass density (middle panel), and temperature (bottom panel) as a function of Rosseland optical depth for model T9. In the upper panel the gray dotted line shows the position of $\tau=2/3$. }
\label{struc_t9}
\end{figure}

Fig.~\ref{struc_t9} illustrates the typical atmospheric structure of the models. The radius at $\tau=2/3$ (where $\tau$ is the Rosseland optical depth) is about six times the radius at the bottom of the atmosphere. Consequently the effective temperature, defined at $\tau=2/3$, is lower than the input temperature: its value is 3606~K.

This difference in the definition of the effective temperature in dense and extended atmospheres is well known for evolved massive stars \citep[e.g.][]{hamann93,hillier98,hillier01}. Because of the thickness of the atmosphere the photosphere is located well beyond the inner radius. This makes comparison between temperatures given by stellar atmosphere and stellar evolution difficult, since the latter usually refer to effective temperatures at their outer boundary, that corresponds to the inner boundary of atmosphere models. When atmosphere are little extended both radii are almost the same. When atmospheres are dense they differ significantly. 

Fig.~\ref{struc_t9} also shows the density structure. Because of the assumption of an expanding atmosphere, the density naturally decreases at lower optical depth. At $\tau=1$, which is characteristic of the value at which the optical continuum forms, the density is of the order $10^{-15}$~g/cm$^3$. The kink structure at an optical depth of about 20 is caused by the rapid increase in velocity at that location. 

The temperature shown in the lower panel reaches $\sim$8000~K near $\tau$=1.0, where most lines are formed.

\section{Formation of \ion{O}{i} lines}
\label{ap_OI}

The formation of the \ion{O}{i} lines at 8446~\AA\ and 11287~\AA\ in LRDs is suspected to be partly driven by Ly$\beta$ fluorescence \citep[e.g.][]{kokorev25}. The upper level of the 8446~\AA\ transition is the lower level of the 11287~\AA\ one. The upper level of \ion{O}{i}~11287~\AA\ is connected to the ground state by a resonance transition that has almost the same wavelength as Ly$\beta$. As a consequence Ly$\beta$ photons can pump the upper level of \ion{O}{i}~11287 which then emits photons by de-excitation. Further cascade down to the lower level of \ion{O}{i}~8446 produce emission in that line too \citep{oliva93}.

\begin{figure}[]
\centering
\includegraphics[width=0.49\textwidth]{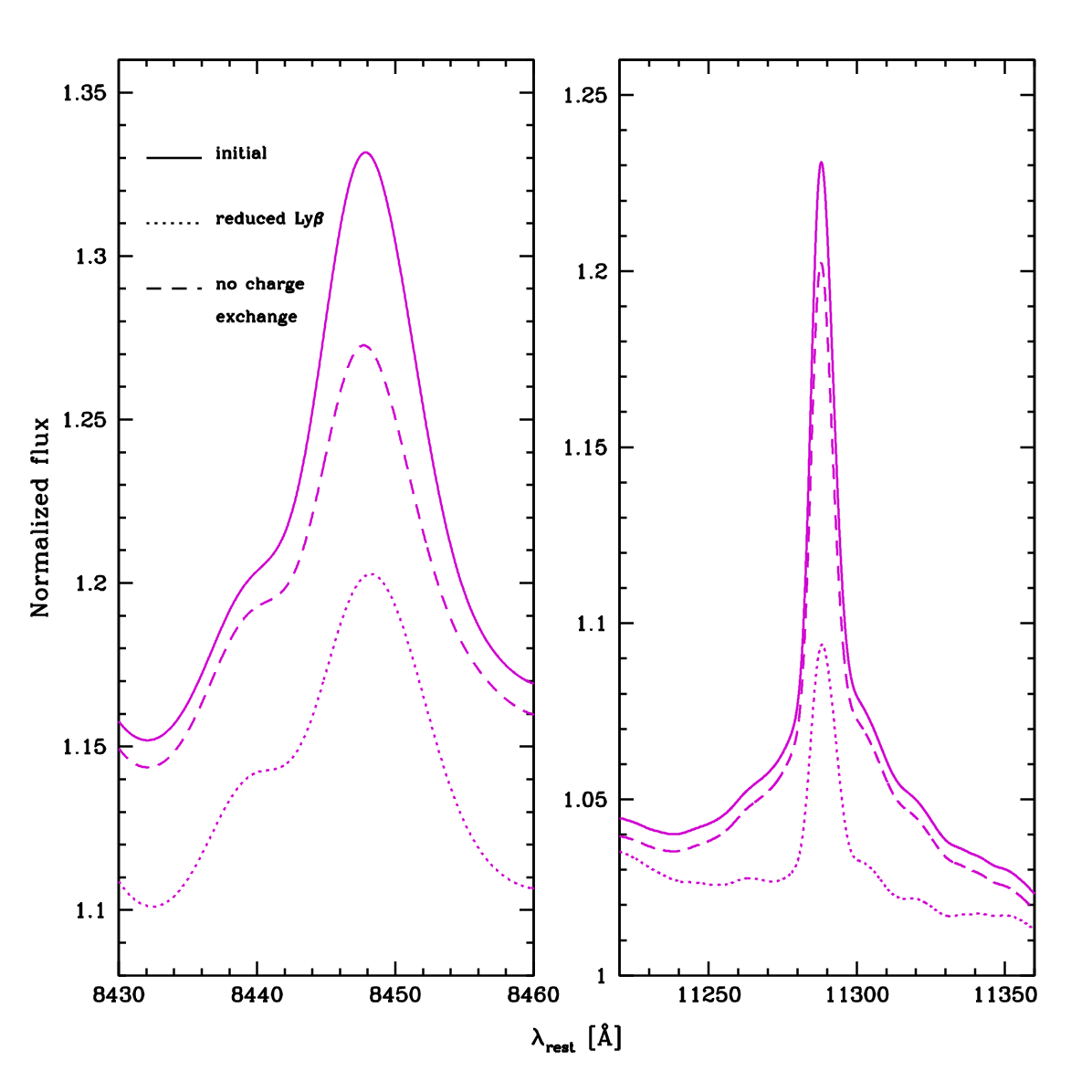}
\caption{Effect of Ly$\beta$ on the intensity of \ion{O}{i}~8446 and \ion{O}{i}~11287. The solid line is the initial model with T=9000~K. The dotted line is the same model in which the oscillator strength of Ly$\beta$ has been reduced by a factor 10$^5$.}
\label{fig_lyb}
\end{figure}

To test if this mechanisms is at work in our models we made the following experiment. Starting from the T6 model at Z~=~0.4~\zsun, we artificially reduced the oscillator strength of Ly$\beta$ by a factor 10$^5$. We converged a new model with only this modification. The result is shown by the dotted line in Fig.~\ref{fig_lyb}. Compared to the initial model (solid line) the \ion{O}{i} lines in this new model are much weaker, confirming the role of Ly$\beta$ in their formation. 

In addition to Ly$\beta$ fluorescence the \ion{O}{i}~8446~\AA\ and \ion{O}{i}~11287~\AA\ lines are potentially affected by charge exchange reactions of the form \ion{O}{ii} + H $\leftrightarrow$ \ion{O}{i} + H$^+$ since the amount of neutral hydrogen is large in the atmospheres studied here \citep[e.g.][]{kokorev25}. Fig.~\ref{fig_lyb} shows how the line profile vary when charge exchange reactions are omitted from the modelling. There is indeed a reduction of the intensity of both lines, but smaller than in the test of Ly$\beta$ fluorescence.

We thus confirm that both effects (charge exchange and fluorescence) are mechanisms that drive the strength of \ion{O}{i}~8446~\AA\ and \ion{O}{i}~11287~\AA. As stressed in Sect.~\ref{s_lines_o} the ratio of the two lines varies with metallicity, indicating that radiative transfer effects are at work. A more detailed study of the formation of \ion{O}{i} lines is postponed to a subsequent publication.

\section{Balmer decrement}
\label{ap_balrat}

In this Appendix we describe the physical conditions of the T6 model to illustrate how a Balmer decrement of the order 1.5 is obtained. In several panels of the figures introduced in this Section we show the contribution function of various lines. This quantity, defined by \citet{hil87}, evaluates the location at which a given amount of energy is emitted in a particular line. It is thus a good indicator of the formation region of lines. 

\begin{figure*}[]
\centering
\includegraphics[width=\textwidth]{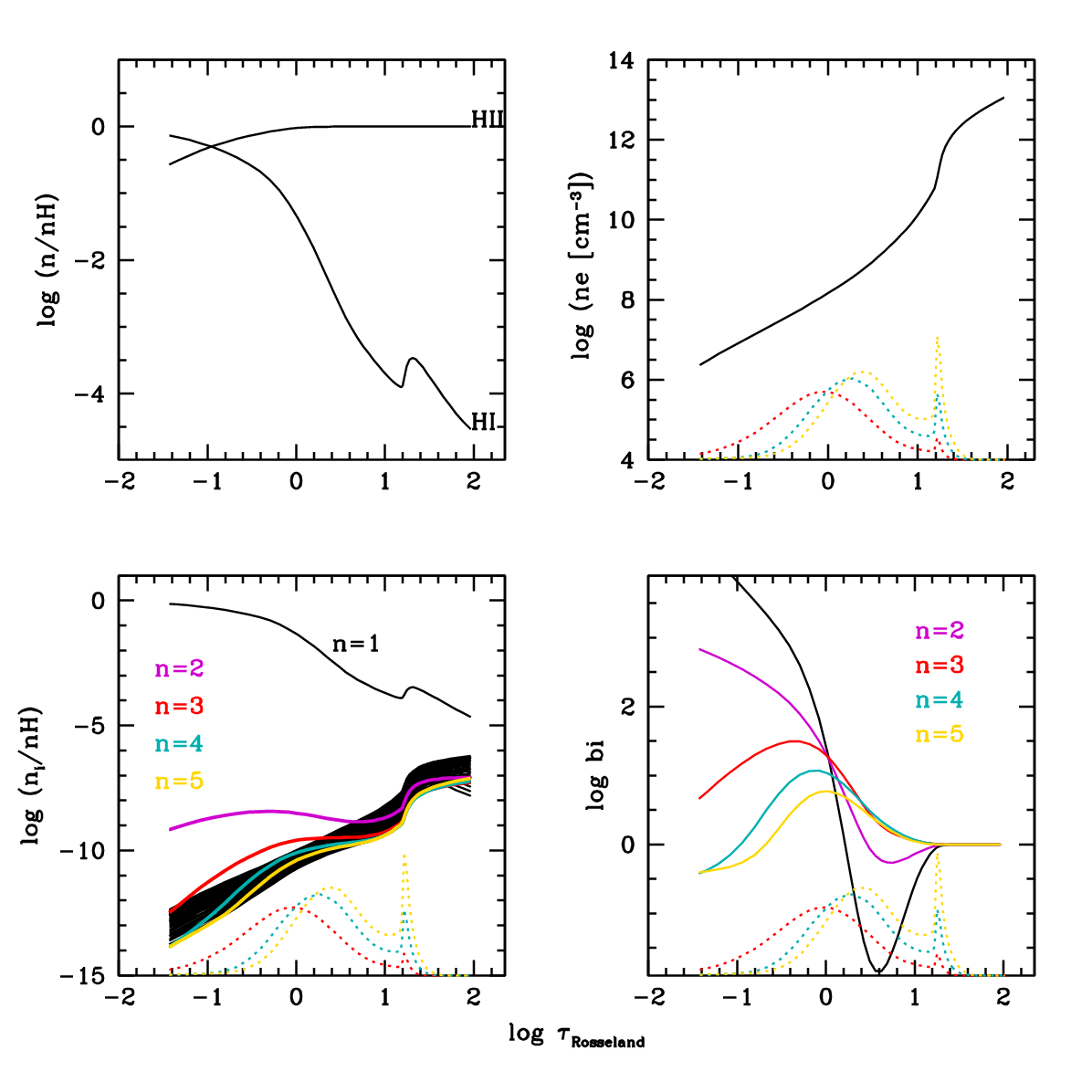}
\caption{Structure of the T6 model. \textit{Top left}: ionization structure, showing the amount of neutral and ionized hydrogen expressed in terms of density divided by the total (neutral + ionized) hydrogen density. \textit{Top right}: electron density. \textit{Bottom left}: relative populations of all hydrogen levels included in the model; the first five levels are highlighted by colors and labels. \textit{Bottom right}: departure coefficient of the first five hydrogen levels. In all panels quantities are plotted against Rosseland optical depth. In the top right and bottom panels the contribution function of \ha, \hb, and \hg\ are shown by the dotted lines.}
\label{fig_popT6}
\end{figure*}

Fig.~\ref{fig_popT6} provides some information on the structure and level populations of the model. From the top left panel one sees that in the inner parts hydrogen is highly ionized. It recombines as one moves to the outer layers where neutral hydrogen becomes the dominant specie. In that model the total \ion{H}{i} column density is $\log$(N(\ion{H}{i}))=23.7. In the top right panel the electron density is shown. It drops from $\log n_{e}$=13 in the interior down to $\log n_{e}$=9 at the maximum of the contribution function of the main Balmer lines. The total electron column density is $\log (N_{e})=25.9$. The bottom panels of Fig.~\ref{fig_popT6} show the level populations and LTE departure coefficients of hydrogen. The latter are defined as the actual population divided by the LTE population evaluated with the local temperature and density. The departure from LTE conditions is striking, with the n=3 to 5 hydrogen level populations being 3 to 8 times higher than LTE populations in the formation regions of \ha, \hb, and \hg.

\begin{figure}[]
\centering
\includegraphics[width=0.49\textwidth]{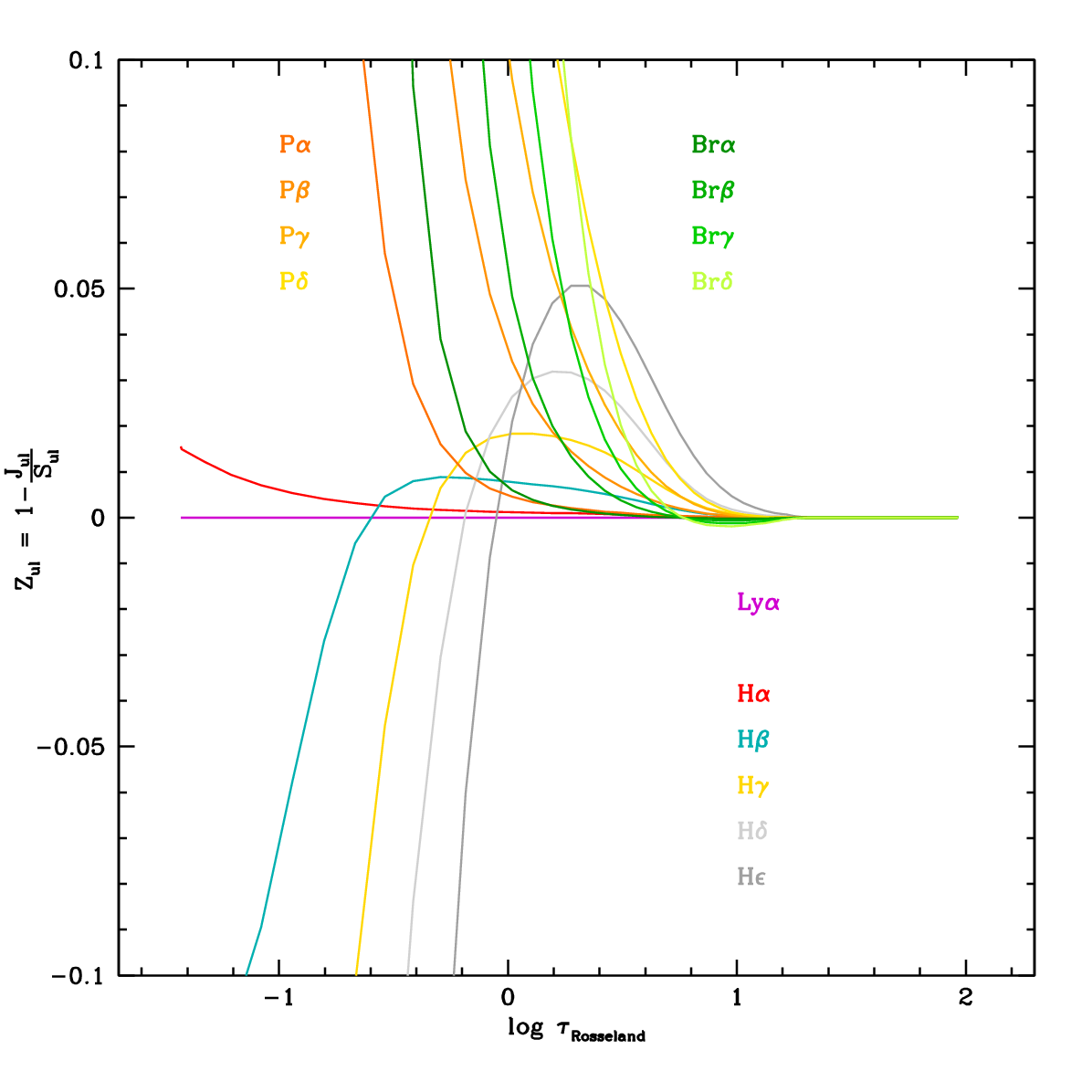}
\caption{Net radiative rates for selected hydrogen lines. }
\label{fig_ZulT6}
\end{figure}

Fig.~\ref{fig_ZulT6} shows the radiative net rates defined as \citep[see][]{mihalas}

\begin{equation}
Z_{ul} = \frac{n_{u} A_{ul} + n_{u} B_{ul} J_{ul} - n_{l} B_{lu} J_{ul}}{n_{u} A_{ul}} = 1-\frac{J_{ul}}{S_{ul}}
\end{equation}

\noindent where $A_{ul}$, $B_{ul}$, and $B_{lu}$ are the Einstein coefficients for transition between the upper and lower levels $u$ and $l$, $J_{ul}$ is the frequency-averaged mean intensity of the radiation field for the transition between the two levels, and $S_{ul}$ the source function for this transition. $Z_{ul}$ evaluates the relative fraction of radiative emission relative to spontaneous emission. A positive $Z_{ul}$ indicates a net emission of photons, while a negative one means a net absorption. A value of zero is encountered in detailed radiative balance. 
In the line formation region of \ha, \hb, and \hg\ (around $\log (\tau_{Rosseland}) = 0-0.5$) the net radiative rate is close to 0 for \ha. It is a factor 5 to 50 larger for \hb\ and \hg. Under nebular case B conditions, Balmer lines radiative rates are equal to 1 since stimulated emission is not taken into account and photo-excitation is not authorized. Consequently the formation of Balmer lines is different in our models, with reduced rates and different values for different lines. Relative to case B, more photons are emitted in \hb\ relative to \ha, which partly explains the lower \ha/\hb\ ratio. 
\citet{drake80} investigated the Balmer decrement under nebular conditions and showed that low values can be obtained under high density conditions, typically $\log n_{e} > 12$, when LTE is almost recovered. 
We have run test models with CLOUDY, exploring the behavior of the Balmer decrement with density. We confirm that the case B value is recovered at low density, then \ha/\hb\ increases up to about 10 for densities of the order $10^{10}$ cm$^{-3}$, and then decreases for higher densities. This confirms the trends of \citet{drake80}.
Our models are clearly not in the high density part of the parameter space as discussed above (lower electron density, strong departure from LTE). We thus attribute the differences in the predicted Balmer decrement to radiative transfer effects. In Fig.~\ref{fig_ZulT6} we see that the net rates behave very differently for different hydrogen series. \lya\ is in detailed balance at all depth. Balmer lines go from net emission to net absorption from the interior towards the exterior. The Paschen and Brackett lines go from detailed balance to strong net emission in the outer regions. This is very different from case B conditions where all rates would be equal to 1 (expect for \lya\ which is by assumption in detailed balance in case B). 

\section{Balmer line asymmetry}
\label{ap_balsym}

\begin{figure}[]
\centering
\includegraphics[width=0.49\textwidth]{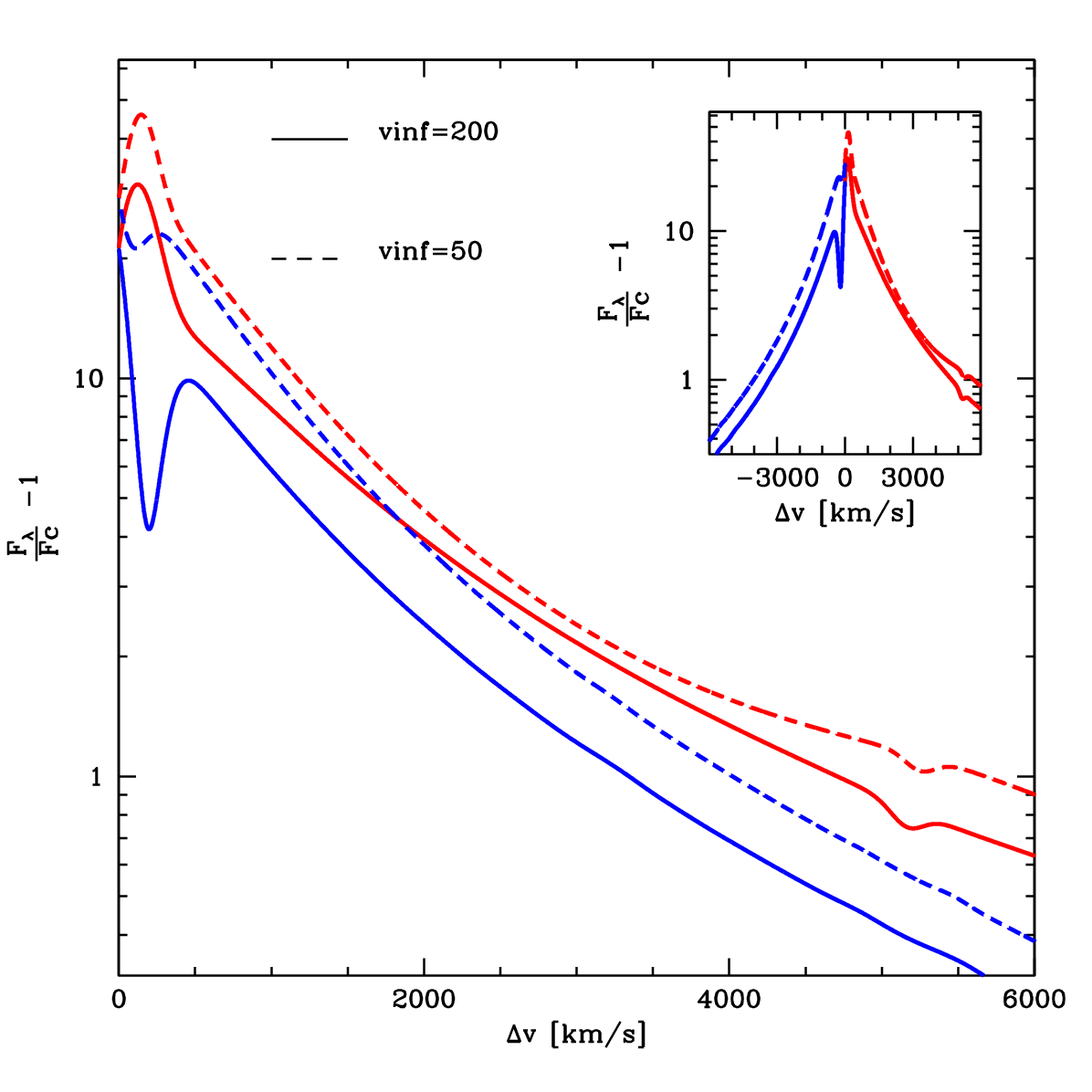}\\
\includegraphics[width=0.49\textwidth]{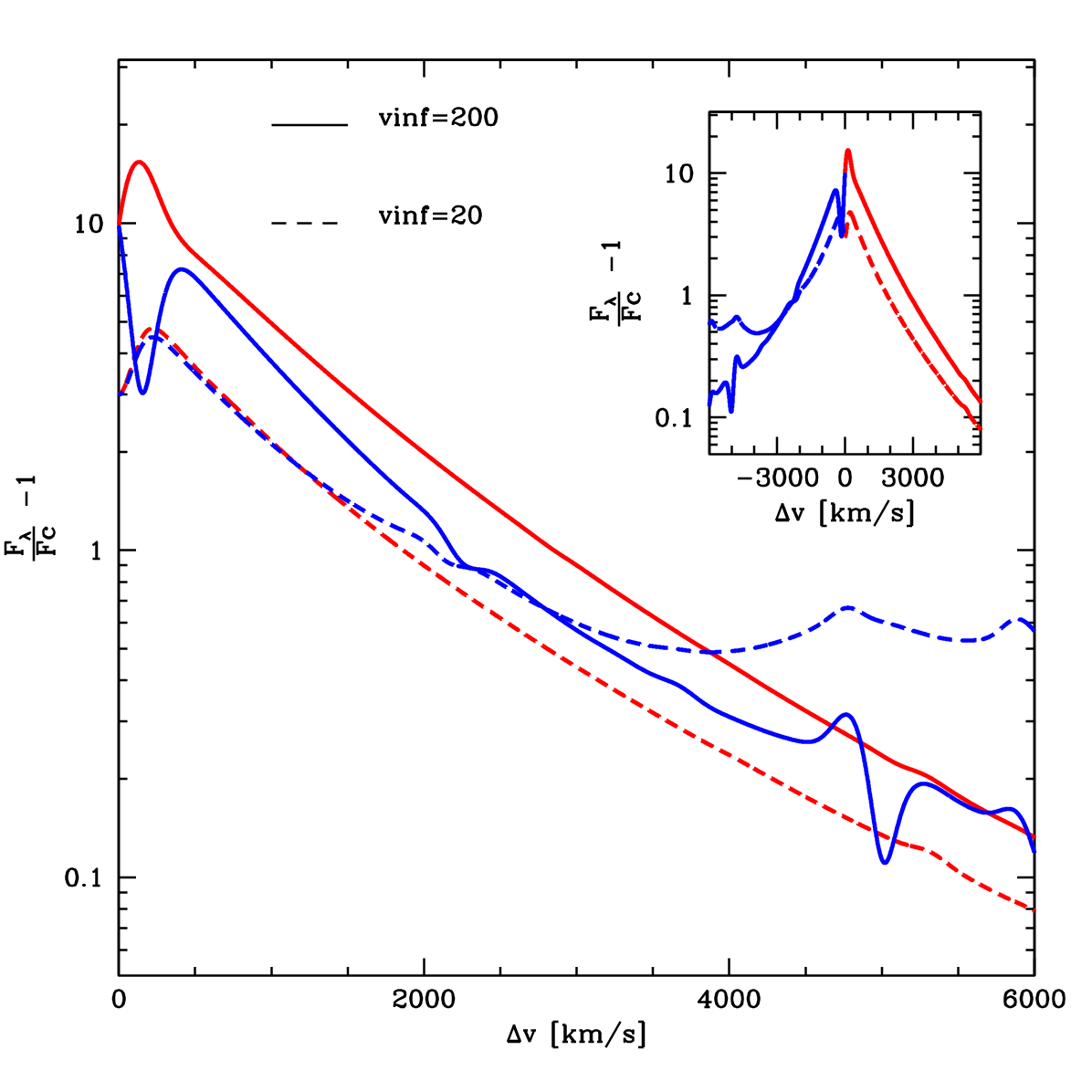}
\caption{Effect of wind terminal velocity (\vinf) on the symmetry of \ha\ for model T12 (top panel) and T6 (bottom panel). The models are homogeneous, i.e. do not include clumping. In both panels the solid line is the initial model, with \vinf\ = 200~\kms. The dashed line is a model in which the velocity is reduced to 50~\kms (top) or 20~\kms\ (bottom). The main panel shows the blue and red wings folded on the same axis, while the insert shows the full profile. Colors are used to highlight the red and blue wings.}
\label{fig_Hasym}
\end{figure}

Fig.~\ref{fig_Hasym} shows the effect of reducing the wind terminal velocity on the (a)symmetry of the \ha\ line profile. The asymmetry of the line is reduced when \vinf\ is reduced. In the right panel the far blue wing is contaminated by line blends, but the central profile is symmetric. In the left panel the line of the small velocity model is not perfectly symmetric and the far red wing shows deviation from symmetry, that can partly be attributed to clumping (see Sect.~\ref{s_lines_h}).

\newpage 

\onecolumn

\section{Effect of atmospheric structure on the SED}
\label{ap_struc}

\begin{figure*}[]
\centering
\includegraphics[width=0.49\textwidth]{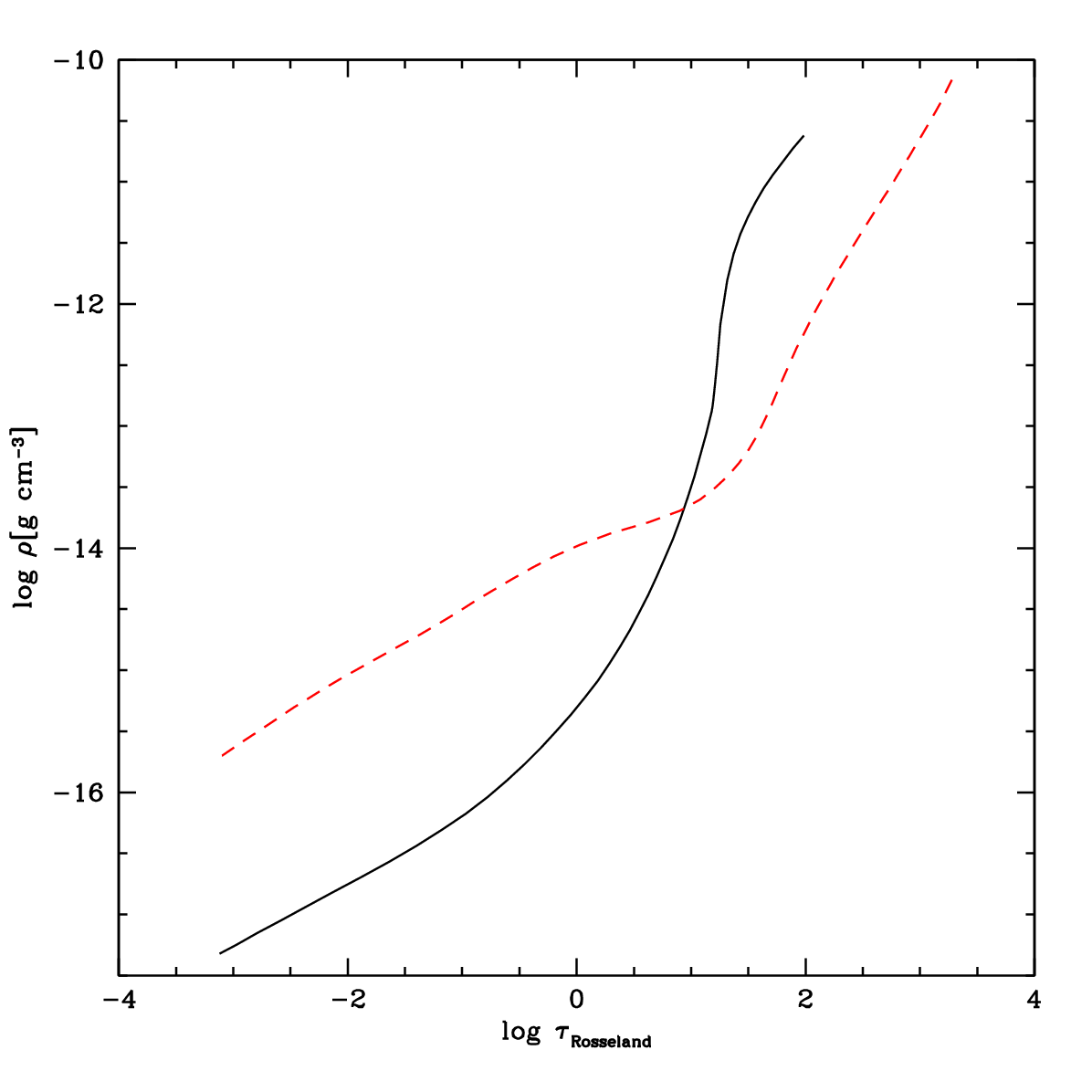}
\includegraphics[width=0.49\textwidth]{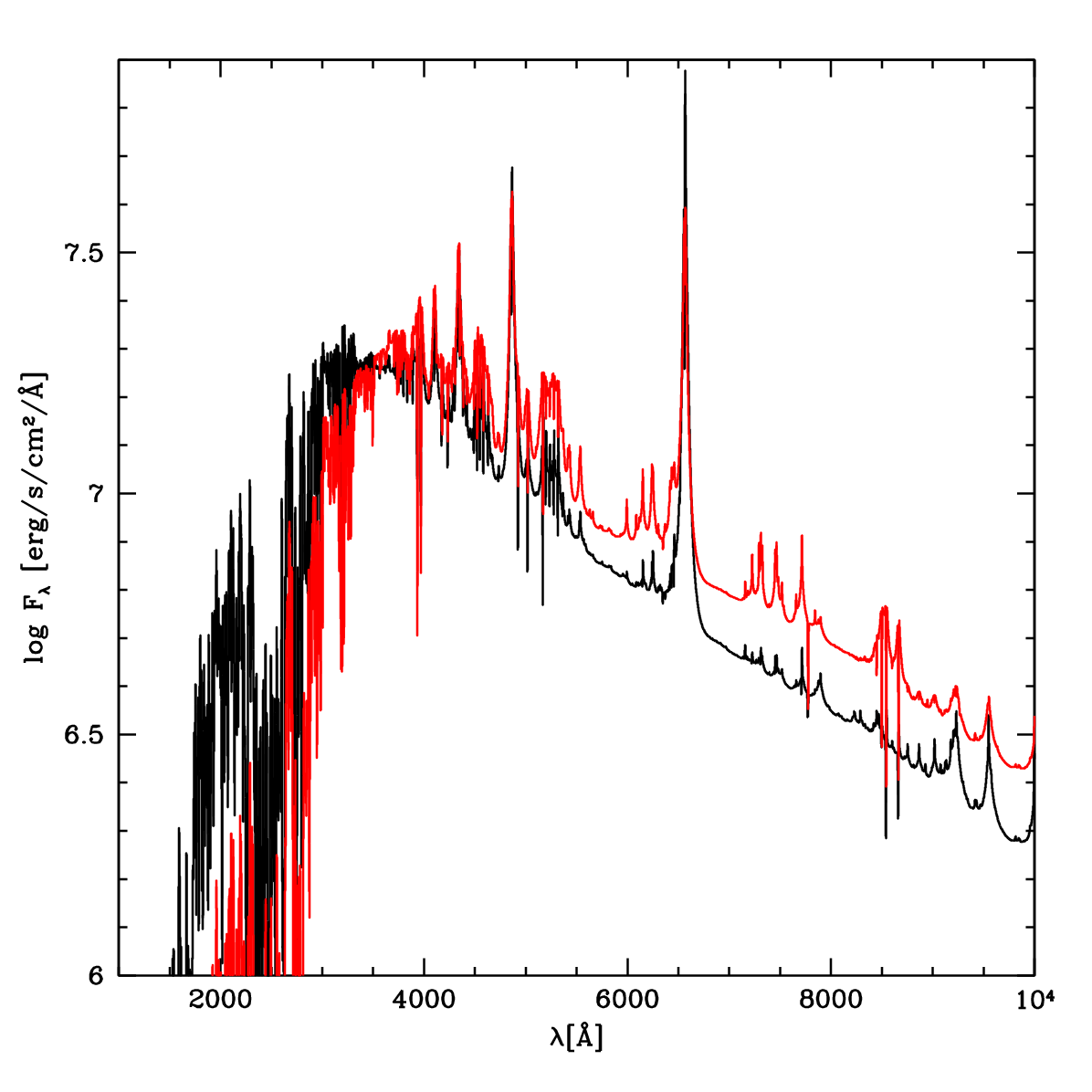}
\caption{Effect of the density structure (left panel) on the emergent spectrum (right panel). The black lines correspond to the T6 model of Fig.~\ref{sed_T}. The red line is a model with the same parameters except the velocity (and thus density) structure which is flatter. The models are for a metallicity Z~=~0.4~\zsun\ which explains the strength of metal lines.
}
\label{struc_T6}
\end{figure*}

Fig.~\ref{struc_T6} illustrates the effect of a change in density structure on the emergent spectrum. The black model is model T6, while the red one is the same model in which the maximum velocity in the atmosphere has been reduced to 20~\kms,  and the scale height increased by a factor 8, thus leading to a flattening of the velocity and density structures. 
 The resulting density structure is shallower than the initial one, but still steeper than a pure $r^{-2}$ structure (that would correspond to a constant velocity structure). There is an effect on the emerging spectrum in the sense that the SED peak is shifted from $\sim$3300~\AA\ to $\sim$3700~\AA. Part of the UV flux is redistributed at longer wavelength since the continuum level of the new model is higher in the optical and near-infrared. The roll-over of the spectrum is thus closer to the Balmer break.

\end{appendix}

\end{document}